# Recognition of potential Covid-19 drug treatments through the study of existing protein-drug and protein-protein structures: an analysis of kinetically active residues

Ognjen Perišić[1]
(1) Big Blue Genomics, Vojvode Brane 32, 11000 Belgrade, Serbia, ognjen.perisic@gmail.com

August 9, 2020

## Abstract

We report the results of our study of approved drugs as potential treatments for COVID-19, based on the application of various bioinformatics predictive methods. The drugs studied include chloroquine, ivermectin, remdesivir, sofosbuvir, boceprevir, and *α*-difluoromethylornithine (DMFO). Our results indicate that these small molecules selectively bind to stable, kinetically active residues and residues adjoining them on the surface of proteins and inside protein pockets and that some prefer hydrophobic over other active sites. Our approach is not restricted to viruses and can facilitate rational drug design, as well as improve our understanding of molecular interactions, in general.

## Introduction

The *Coronaviridae* positive stranded RNA virus family includes a substantial number of members, many of whom are known to cause a broad range of illnesses from common cold to sever diseases like Severe Acute Respiratory Syndrome (SARS), Middle East Respiratory Syndrome (MERS), etc. [1, 2]. The latest worldwide rapidly spreading disease, COVID-19, is caused by a new member of this virus family, SARS-COV-2. The disease originally emerged in China in December 2019 with most common symptoms being fever and cough, as well as shortness of breath, sore throat, headache, muscles ache, nausea, and diarrhea [1]. In some cases symptoms also involve a still unexplained loss of smell and taste [3]. It is assumed that the SARS-COV-2 virus spreads through respiratory droplets, directly via physical contact, or through contact with contaminated objects, and can severely affect patients with immune systems weakened by pre-existing conditions, such as hypertension, diabetes mellitus or cardiovascular diseases [4]. The virus spreads more easily than SARS and MERS due to high binding affinity between the virus spike glycoprotein (S) and the host receptor [5-8], making it more deadly. By forcing countries to restrict access to work and thus slowing down supply lines the virus directly affects the global economy, which experiences a significant decline in gross national products worldwide not encountered since the Great depression.

There are a number of efforts and clinical trials underway to develop a vaccine and evaluate potential drugs for COVID-19, but such investigations usually take many months or even years to yield a successful treatment. Drug repurposing, on the other hand, may offer an immediate solution, because it considers already approved compounds as potential treatments for COVID-19. There are two paths toward a viral treatment. One path directly attacks the virus and interrupts its replication machinery or its ability to attack host cells [9]. This path is often hard to implement due to rapid emergence of new viral strains with acquired resistance to implemented drugs. The second path should therefore aim to block the host-viral interactions on the host side due to difficulties viral single point mutations should have in recovering the loss of host factors [10]. A recent study of human-virus protein-protein interactions (PPIs) detected 332 high-confidence SARS-CoV-2-human PPIs [11]. The study showed that 40% of SARS-CoV-2 proteins interact with endomembrane compartments or vesicle trafficking pathways, and that viral proteins also interact with multiple innate immune pathways, the host translation machinery, bromodomain proteins, enzymes involved in ubiquitination regulation, and Cullin ubiquitin ligase complex. Importantly, it showed that the SARS-COV-2 human PPI map is very similar to the interaction maps of West Nile Virus (WNV) and Mycobacterium tuberculosis (Mtb). Among the human proteins involved in interactions with viral proteins, the study detected 66 druggable human (host) proteins targeted by 69 compounds (29 FDA-approved drugs, 12 drugs in clinical trials, and 28 preclinical compounds). It identified two groups of compounds with noticeable antiviral activity: inhibitors of mRNA translation/protein biogenesis (zotatifin, ternatin-4, PS3061, and plitidepsin), and predicted regulators of the Sigma1 and Sigma2 receptors (Haloperidol, PB28, PD-144418, Hydroxychloroquine, Clemastine, Cloperastine, Progesterone, and the clinical molecule Siramesine). The first group of compounds directly affects the viral cap-dependent mRNA translation because coronaviruses use the host translation machinery for their own mRNA translation. The compounds affecting the second group of proteins are approved and long established human therapeutics [11]. As much as they are informative, such screening associative studies rarely offer detailed insights into mechanisms of molecular interactions, whereas structural studies [5-8, 12] give snapshots into residue and atom level physical interactions between molecules, but cannot offer general principles of molecular interactions.

In-silico studies are widely used to screen potential drug candidates against Covid-19. Recently, Calligari et al. reported a molecular docking study binding affinities between different viral proteins and several inhibitors, originally developed for other viral infections (HCV, HIV) [13]. The examined drugs include Simeprevir, Saquinavir, Indinavir, Tipranavir, Faldaprevir, Ritonavir, Lopinavir, Asunaprevir, Atazanavir, Nelfinavir, Amprenavir, Darunavir and Fosamprenavir. They docked them with the 3C-like protease from SARS–CoV-2, the spike S SARS-Cov-2 protein, SARS-Cov-2 RdRp, and nucleocapsid protein from SARS-Cov-2, and reported the crorresponding binding free energies. The authors were able to select 5 of the 13 as potential inhibitors of SARS-Cov-2 protease. The three drugs tested with the Spike S protein showed promise and one presumably binds between the two monomers and interrupts transition between opened and closed states. A.A. Elfiky used molecular modeling, docking, and dynamics simulations to build a model for the viral RdRp proteins and test its binding affinity to some clinically approved drugs and drug candidates [14]. The author concludes that sofosbuvir, ribavirin, galidesivir, remdesivir, favipiravir, cefuroxime, tenofovir, and hydroxychloroquine can tightly bind to the RdRp active site and can be good candidates for clinical trials. The author also noticed that the compounds setrobuvir, YAK, and IDX-184 can tightly wrap to the SARS-CoV-2 RdRp, and thus interrupt its function. That study also showed the IDX-184 derived compounds (3,5-dihydroxyphenyl)oxidanyl and (3-hydroxyphenyl)oxidanyl can be effectively used to target SARS-CoV-2 RdRp. Yu et al. used AutoDock Vina software to screen potential drugs by molecular docking with the structural protein and non-structural protein sites of Covid-19 virus [15]. They tested ribavirin, remdesivir, chloroquine, and luteolin, a compounds present in Honeysuckle. In traditional Chinese medicine honeysuckle is generally believed to have antiviral effects. In this study, luteolin (the main flavonoid in honeysuckle) was found to bind with a high affinity to the same sites of the main protease of SARS-CoV-2 as the control molecule. De Oliveira et al. tested 9091 drug candidates by molecular docking against equilibrated SARS-Cov-2 Spike S protein [16]. 24 best-scored ligands, ivermectin among them, 14 of them are traditional herbal isolates and 10 are approved drugs, exhibited binding energies below -8.1 kcal/mol, and were thus suggested as potential candidates. O. Santos-Filho used molecular docking to test HIV protease inhibitors against Covid-19 main protease [17]. The author's study showed that two non-natural compounds, danoprevir and lopinavir, and one herbal compound, corilagin, produced strong interactions with the inhibitor binding site of SARS-CoV-2 main protease. A modeling study by Pachetti et al. recognized a number of Covid-19 RdRp mutations that can affect drug treatments against Covid-19 RdRp [18]. B. R Beck et al. used pre-trained deep neural network to identify commercially available drugs that could be used as treatments against SARS-CoV-2 [19]. They showed that atazanavir, an antiretroviral medication used to treat and prevent the human immunodeficiency virus (HIV), is the best compound against the SARS-CoV-2 3C-like proteinase, followed by remdesivir, efavirenz, ritonavir, dolutegravir, lopinavir and darunavir.

To facilitate the drug screening we undertook a comparative in-silico study of binding modes of six antiviral candidate drugs. We analyzed compounds that bind to parasitic and to human proteins. We use those results to anticipate their binging affinities. The drugs we have studied so far include (hydroxyl)chloroquine, ivermectin, remdesivir, sofosbuvir, boceprevir, and $\alpha$-difluoromethylornithine (DMFO) (see Table 1).

**Chloroquine** [20] and its less toxic derivative **hydroxychloroquine** [21] are drugs used to prevent and treat acute attacks of malaria. They are also used to treat discoid or systemic lupus erythematosus and rheumatoid arthritis in patients whose symptoms have not improved with other treatments. These drugs are subject of a number of clinical trials worldwide as potential treatment for Covid-19 [22, 23]. Interestingly, the above mentioned study [11] shows that PB28 is ~20 times more potent viral inhibitor than hydroxychloroquine.

**Ivermectin** is a medication used to treat many various types of parasite infestations [24]. They include, but are not limited to, head lice, scabies, river blindness (*onchocerciasis*), *strongyloidiasis*, *trichuriasis*, *ascariasis*, and *lymphatic filariasis*. Depending on the kind of treatment, the drug is taken by mouth or applied to the skin for external infestations. Ivermectin molecular structure is rather complex and made of a set of macrocyclic lactone isomers. Ivermectin binds to glutamate-gated chloride channels and increases the permeability of chloride ions. The drug was shown to inhibit the replication of SARS-COV-2 in vitro [25] and is currently the subject of clinical trials as a potential COVID-19 treatment [26].

**Remdesivir** is a nucleoside analog RNA-dependent RNA Polymerase (RdRp) inhibitor initially developed to treat Ebola and Marburg virus diseases [9, 27]. The drug decreases the viral RNA production by affecting the function of RdRp and proofreading by viral exoribonuclease (ExoN). Remdesivir is a subject of clinical trials as a potential COVID-19 treatment [28], as it was shown to reduce the lung viral load and improve pulmonary function with SARS infection [9].

**Sofosbuvir** is a medication used to treat HCV mono-infection and HCV/HIV-1 coinfection as component of a combination antiviral regimen [29]. Sofosbuvir is nucleotide prodrug that metabolically gets modified to the active uridine analog triphosphate, an inhibitor of HCV NS5B RNA-dependent polymerase. The inhibition of HCV NS5B RNA-dependent polymerase in turn suppresses viral replication. A. Sadeghi presented a tentative results on the effectiveness of sofosbuvir and daclatasvir agains Covid-19 [30].

**Boceprevir** is a medication used to treat chronic hepatitis C in untreated people or who do not react to ribavirin and peginterferon alfa alone [31]. It is used in combination with ribavirin (Copegus, Rebetol) and peginterferon alfa (Pegasys). It was shown to inhibit the Covid-19 (SARS-Cov-2) replication by inhibiting the virus main protease [32].

**α-difluoromethylornithine (DMFO)** (Eflornithine), is a medication primary used to treat *African trypanosomiasis* (sleeping sickness) and excessive facial hair in women [33]. Specifically, it is used for the second stage of sleeping sickness caused by *T. b. gambiense* and may be used with nifurtimox [34]. It is used by injection or applied to the skin. The drug prevents binding of the natural product ornithine to the active site of ornithine decarboxylase. We did not find any record of this drug ever being tried, so far, for COVID-19. However, since it is a halogenated organic molecule with somehow similar active sites as Chloroquine we decided to study it towards treatment of COVID-19.

Here we report our research findings based on the method which we have implemented to recognize protein-protein binding patterns, the Self-Adjustable Gaussian Network Model (SAGNM) [35], and on the existing bioinformatics predictive methods and tools, Chimera and VMD [36, 37]. Our SAGNM method predicts binding areas without any information on the binding partner's properties, position or orientation.

The analysis of drug binding spots on the surfaces of protein targets is based on the Gaussian Network Model formalism [38-43]. The GNM produces a set of vibrational modes via the eigenvalues and eigenvectors of the protein Kirchhoff contact matrix $\Gamma$. The fastest modes (with largest eigenvalues $\lambda$) are more localized and have steeper energy walls with a larger decrease in entropy and they are, therefore, referred to as kinetically hot residues. For more details on Phantom network and GNM, see the Supplementary Materials in [35].

The connection between kinetically hot residues and interfacial residues involved in protein-protein interactions has already been established [44]. The methodology introduced in [35], and used here is based on a self-adjusting approach. It is able to accurately and effectively determine binding pockets for peptides and small, drug-like molecules.

The term "kinetically hot residues" is linguistically close, but does not carry the same meaning as the term "hot spots" that is often used in protein science. Hot spots are residues that often appear in structurally preserved interfaces (in more than 50% of cases). They are important, because they are general contributors to the binding free energy. They are defined as spots where alanine mutation increases the binding free energy at least 2.0 kcal/mol [45-51].

## Methods and Materials

Our aim was to analyze presently available structures existing drugs bound to parasitic and human proteins and predict their binding patterns, as well as the binding patterns of SARS and SARS-COV-2 binding patterns to the human ACE2 receptor. To predict binding residues in protein we applied our Self Adjustable interpretationof Gaussian Network Model (SAGNM) [35]. The structure alignment, hydrophobicity calculation, and visualization and analyses were performed with the programs Chimera [36] and VMD [37].

The software for the Self Adjustable Gaussian Network Model code is composed of several different programs. The first program calculates contact maps and the corresponding eigenvectors and eigenvalues [52] for both protein chains forming a protein dimer (given as a PDB file). To accomplish that the program first calculates the Kirchhoff contact matrix $\Gamma$ for each protein. The matrix $\Gamma$ calculation is based on the distances between $C_\alpha$ atoms only, and those distances have to be lesser or equal to 7 Å to consider two residues in a contact [38-40]. The code then calculates and sorts $\Gamma$ matrix eigenvalues and eigenvectors. The eigenvectors are sorted according to their corresponding eigenvalues. Those eigenvalues and eigenvectors are used in the second part that (iteratively) calculates the weighted sum of modes [41] as

$$\left\langle (\Delta \mathbf{R}_i)^2 \right\rangle_{k_1-k_2} = (3k_B T/\gamma) \sum_{k_1}^{k_2} \lambda_k^{-1} [\mathbf{u}_k]_i^2 \Bigg/ \sum_{k_1}^{k_2} \lambda_k^{-1}, \qquad (1)$$

where $\lambda_k$ are eigenvalues and $\mathbf{u}_k$ are eigenvectors. See the Supplementary material in [35] for details on phantom network and GNM.

This equation, normalized by dividing the sum by $(3k_B T/\gamma)$ produces mean square fluctuations of each residue by a given set of modes ($k1$ to $k2$). The equation produces an estimate of a kinetic contribution of each residue for that set of modes. The above equation is very similar to the singular value decomposition method [53] used in the linear least squares optimization method. An additional code extracts contact and first layer residues. Finally, the third set of routines extracts neighboring residues and their distances for each residue per protein chain. That information is later used in the spatial spreading of the influence of kinetically hot residues.

The first step in the SAGNM procedure is the calculation of the weighted sum (Eq. 1). The procedure starts with a number of modes that corresponds to the top 10% of eigenvalues range of the analyzed protein. With normalized sum only residues with a normalized amplitude higher than 0.05 are perceived as hot residues. The number of hot residues is usually smaller than the number of potential contact or first layer residues (not contact residues with a spatial atom-atom distance of less than 4.5 Å from contact residues). To account for that the influence of hot residues is spread to their sequential neighbors using the sequence information obtained from protein structure PDB files (to account for possible missing residues). The influence of hot residues is first spread linearly, to sequential neighbors only, because proteins are polymer chains with physically connected residues. That implies that sequentially neighboring residues should exhibit correlated behavior. For chains longer than 100 amino acids (aa), hot residues and 8 their sequential neighbors upstream and downstream are labeled as predictions (four upstream, four downstream). For shorter chains the influence is spread to 6 neighboring residues.

The prediction is then expanded to spatial neighbors. This approach is much closer to the true nature of the GNM algorithm that uses only spatial distances between $C_\alpha$ atoms and disregards any sequential/connectivity information. To apply this approach, the maximum cutoff $C_\alpha$-$C_\alpha$ distance from the center of a hot residue was introduced to which its influence can be spread. The cutoff distance of 6 Å was applied with for shorter protein chains (for sequence lengths shorter than 250 aa) and the cutoff of 8 Å for longer protein chains. All residues with $C_\alpha$ atoms within the sphere centered at the $C_\alpha$ atom of the hot residue, and within the assigned cutoff distance, are considered to be predictions. The two cutoff values were estimated empirically [35]. To extract spatial neighbors, distances between residues ($C_\alpha$-$C_\alpha$ distances) were calculated for each particular protein and sorted in ascending order.

The Self Adjustment GNM scheme is performed as follows:

**Step 1:** Calculate the number of fast modes that correspond to the top 10% of eigenvalues range.
**Step 2:** Calculate the weighted sum (Eq. 1) and spread the influence of hot residues to sequential and spatial neighbors.

**Step 3a:** If the overall percent of predictions is larger than expected (for example, if the percent of predictions is larger than the previously set value of the 30% of the total number of residues), the SAGNM procedure reduces the number of fast modes by one and goes to the **Step 2**.
**Step 3b:** If the percent of predictions is too small (e.g. less than the predetermined value of 15% of all residues), the SAGNM procedure increases the number of fast modes by one and goes to the **Step 2**.

The Self Adjustable procedure repeats the **Steps 2** and **3** until the percent of predictions fits between the maximum and minimum expected percentages for a given chain.

To avoid infinite loops, only one increase followed by a decrease is allowed, and vice versa. Multiple consecutive increases or decreases are allowed. This approach ensures that longer proteins have enough predictions, and that shorter ones are not saturated with too many false positives.

We focused our study on pdb structures with the listed drugs present as ligands. For chloroquine we analyzed 2 structures (pdb ids 1cet [54] and 4v2o [55]). For Ivermectin we analyzed the binding pattern of the drug to the human glycine receptor alpha-3 (pdb id 5vdh [56]). For Remdesivir we analyzed its binding patterns in the recently released structure (pdb id 7bv2 [12]). We also performed the analysis of binding pattern of drug Sofosbuvir to the hepatitis C virus (HCV) RdRp (pdb id 4wtg [57]) and compared them to the Covid-19 RdRp predictions (pdb id 6m71 [58]). Sofosbuvir was already analyzed in light of similarities between HCV and SARS-COV-2 RdRp and similarities between Remdesivir's and Sofosbuvir's [59]. For boceprevir we analyzed the structure SARS-Cov-2 main protease bound to the drug (pdb id 6wnp). For α-Difluoromethylornithine we analyzed a structure of *Trypanosoma brucei* ornithine decarboxylase (ODC) with D-ornithine bound to it (pdb id 1njj [60]). α-Difluoromethylornithine binds to the active site of ODC and inhibits ornithine binding to it. We performed the comparative analysis of the binding patterns between the ACE2 human receptor and the spike glycoproteins from SARS (pdb id 6cs2 [61]) and SARS-COV-2 (pdb id 6m0j [7]). We also analyzed the binding patterns between the SARS RBD with S230 human neutralizing antibody, and between SARS RBD and glycan shield (pdb id 6nb6 [62]).

# Results

## Chloroquine
The analysis of chloroquine binding patterns to *Plasmodium Falciparum* Lactate Dehydrogenase (pdb id 1cet [54]) and human Lysosomal protein Saposin B (pdb id 4v2o [55]) reveals that chloroquine binds to kinetically active sites recognized by the SAGNM algorithm which are mostly hydrophobic (Figure 1).
With Lactate Dehydrogenase chloroquine binds selectively and competitively to the Nicotinamide adenine dinucleotide (NADH) binding pocket of the enzyme, and occupies a position similar to that of the adenyl ring of the cofactor. It is thus a competitive inhibitor for this critical glycolytic enzyme of malaria [54]. The SAGNM algorithm recognizes residues Val-24, Leu-25, Val-48, Leu-51, Ala-63 and Val-94 as hot. Their influence is spread to the residues Lys-20, Ala-21, Lys-22, Ile-23, Val-24, Leu-25, Val-26, Gly-27, Ser-28, Gly-29, Gly-32, Ala-37, Ile-40, Asn-44, Leu-45, Gly-46, Asp-47, Val-48, Val-49, Leu-51, Phe-52, Asp-53, Ile-54, Val-55, Pro-59, His-60, Gly-61, Lys-62, Ala-63, Leu-64, Asp-65, Thr-66, Ser-67, Cys-76, Lys-77, Val-78, Ser-79, Gly-80, Ser-81, Asp-87, Leu-88, Gly-90, Ser-91, Asp-92, Val-93, Val-94, Ile-95, Val-96, Thr-97, Ala-98, Ala-133,

Phe-134, Ile-135, Ile-136. See Figure 1a-d and Figure S1a in the Supplementary material. The drug interacts with residues Val-26, Gly-27, Phe-52, Asp-53, Ile-54, Tyr-85, Ala-98, Phe-100, Ile-119 and Glu-122 and the SAGNM algorithm correctly recognized residues 26, 27, 52, 53, 54 and 98. Other sites although exposed to solvent are not binding targets. The Chloroquine molecule binds preferentially to hydrophobic sites (see Figs 1c-d) and avoids neutral and hydrophilic areas. The lysosomal protein Saposin B is a trimer formed by chains A, B and C. It selectively degrades lipids and is one of the most studied members of the saposin protein family [55]. Its deficiency or malfunctioning leads to the accumulation of lipids in the lysosome and results in the lysosomal storage disease metachromatic leukodystrophy ([55] and references therein). The SAGNM algorithm recognizes the residues Ile-8 and Cys-71 in chain A, Ile-8 and Cys-71 in chain B and Cys-71 in chain C as hot. Their influence is spread to the residues Gln-5, Asp-6, Cys-7, Ile-8, Gln-9, Met-10, Val-11, Pro-67, Lys-68, Glu-69, Ile-70, Cys-71, Ala-72, Leu-73, Val-74, Phe-76 and Cys-77 in chain A; to the residues Gln-5, Asp-6, Cys-7, Ile-8, Gln-9, Met-10, Val-11, Pro-67, Lys-68, Glu-69, Ile-70, Cys-71, Ala-72, Leu-73, Val-74, Phe-76 and Cys-77 in chain B, and to the residues Lys-68, Glu-69, Ile-70, Cys-71, Ala-72, Leu-73, Val-74, Phe-76 and Cys-77 in chain C. See Figure 1e-h and Figure S1b in the Supplementary material. The SAGNM recognizes that the chloroquine molecules interact with residues Glu-69 and Leu-73 from chain B, out of residues Ala-58, Met-61, Met-65, Glu-69 and Leu-73. With chain C, it recognizes residues Glu-69 and Leu-73, out of residues Met-61, His-64, Met-65, Glu-69 and Leu-73. With chain A it does not emphasize the residue Arg-38, but it recognizes the binding patch with the chain C (see Figure 1e).

With both chloroquine examples the expected number of predictions for the SAGNM algorithm was set to be between 10% and 15%, and that corresponds to the fastest normal mode. Our results suggest that chloroquine's binding to Covid-19 proteins should follow the same patterns as with Lactate Dehydrogenase and Saposin B. Namely it should attach to residues which are both hydrophobic and kinetically active (or very close to kinetically active sites).

The analysis of chloroquine's nondiscriminatory binding to human and parasitic proteins may offer an explanation of its efficiency against parasitic infections as well offer a glimpse into its toxicity.

## Ivermectin

The drug ivermectin binds glutamate-gated chloride channels and thus increases their permeability to chloride ions. We analyzed the ivermectin's binding to the human glycine receptor alpha-3 (pdb id 5vdh [56]). This structure, besides ivermectin, also has glycine and the potentiator AM-3607 (7c6) bound to the glycine receptor. The comparison of the crystal structure used in this research to previously determined structures revealed that the ivermectin binding expands the ion channel pore [56].

The receptor is a pentamer, so we only analyzed the binding to its chain A (Figure 2). The SAGNM algorithm recognized the residues Glu-157, Ser-158, Phe-168, Phe-207, Thr-208, Cys-209, Ile-210, Glu-211, Ser-238, Gly-256, Thr-259, Val-260, Val-294 and Leu-298 as kinetically hot. For the list of residues their influence is spread to see the list below Figure S2 in the Supplementary material. The expected number of predictions for the SAGNM algorithm was set to be between 25% and 30%, and that corresponds to 3 fastest modes. The three compounds bind to the residues Arg-27, Ile-28, Arg-29, Phe-32, Phe-159, Gly-160, Tyr-161, Asp-165, Tyr-202, Thr-204, Phe-207, Ser-267, Ser-268, Ser-278, Val-280, Asp-284, Ala-288, Leu-291, Leu-292 and Phe-295. The SAGNM algorithm recognized residues Phe-159, Gly-160, Tyr-161, Asp-165, Tyr-204, Phe-207, Leu-291, Leu-292 and Phe-295. The analysis (Figure 2) reveals that all three compounds (ivermectin, glycine and AM-3607 (7c6)) bind to kinetically active and adjoining residues [35], some of which are highly hydrophobic, with ivermectin binding almost exclusively hydrophobic residues. That means that this drug well seek similar sites on the surface of the Covid-19 proteins.

# Remdesivir

We used the recently cryo-EM determined structure of SARS-COV-2 RdRp with double-stranded template-primer RNA and remdesivir (pdb id 7bv2 [12]) to analyze the RNA and drug binding to residues in RdRp. The structure reveals that the double stranded RNA is inserted into RdRp's central channel and that the active triphosphate form of remdesivir is covalently bound to the primer strand at the first replicated base, which effectively terminated the chain elongation. It should be noted that the prodrug form of remdesivir does not have any inhibitory effect on the polymerization activity of the purified enzyme [12]. The SAGNM algorithm recognized the residues Gly-503, Thr-538, Ile-539, Thr-540, Gln-541, Ala-558, Val-560, Ser-561, Val-609, His-613, Glu-665, Met-666, Val-667, Met-668, Ala-702, Ala-706, Phe-753, Cys-765 and Asn-767 in chain A as hot, the residues Asp-161 and Ile-185 in chain B as hot, and the residues Lys-7, Ser-10, His-36, Ile-39, Ala-48 and Lys-51 in chain C as hot. For the predictions see the list below Figure S3 in the Supplementary materials.

Our analysis reveals that the residues recognized via the fastest two normal modes (corresponding to kinetically active residues [35]) delineate the central channel (Figure 3a-b). The enzymatically important residues Lys-500, Ser-501, Lys-545 and Arg-555 are all recognized by the SAGNM algorithm using just the fastest normal mode, while the residue Asp-761 of the catalytic center (out of residues Ser-759, Asp-760 and Asp-761 that form the catalytic center) is also emphasized with the two fastest modes. Residues Lys545 and Arg-555 are important because they stabilize the incoming nucleotide in the correct position for catalysis. The crystal structure shows that the catalytic center of RdRp, NSP12 protein (Non Structural Protein 12), does not have any contacts with base pairs of RNA emphasizing RdRp's sequence-agnostic polymerization ability [12]. This is in concordance with our coarse grained analysis, based on the positions of C-α atoms only, that shows that stable, kinetically active residues outline the enzyme's central channel.

# Sofosbuvir

We performed a comparative analysis of the Hepatitis C virus (HCV) RdRp (chain A in pdb id 4wtg [57]; the structure is given with the drug sofosbuvir bound to it) and the Covid-19 RdRp (chain A in pdb id 6m71 [58]). We followed the steps of Y. Gao and collaborators [59] and attempted to compare predictions of the binding residues in HCV RdRp to sofosbuvir, to binding residues predictions in Covid-19 RdRp. The binding residues in HCV are buried deep inside the polymerase catalytic core. Our analysis shows that they are generally delineated by the kinetically active residues and are thus stable and characterized by the two fastest normal modes (Figure 4a), but they are not explicitly hydrophobic (Fig. 4b-c). The SAGNM algorithm recognized the residues Met-139, Ala-157, Met-266, Asn-268, Cys-279, Lys-298, Phe-339 and Met-343 of the HCV RdRp (pdb id 4wtg) as hot. The algorithm recognized the residues Met-139, Ala-157, Met-266, Asn-268, Cys-279, Lys-298, Phe-339 and Met-343 of the main enzymatic unit of Covid-19 RdRp (pdb id 6m71) as hot. For the full list of hot residues and predictions for HCV RdRp see Figure S4 and the list below it, and for Covid-19 RdRp see Figure S5 and the list below it in the Supplementary material.

The structural alignment of HCV and Covid-19 RdRp (Figure 4e) using the Chimera program [36] shows that they share the structure of the binding pocket, and also reveals that the catalytic cores in both proteins is bounded by kinetically active residues, but the overall distribution of residues is only partially similar between the two proteins (Figure 4f). In both cases the expected number of targets is between 15 and 20%. With HCV RdRp that corresponds to the two fastest modes, and with Covid-19 RdRp and the main catalytic unit NSP12, that corresponds to 7 fastest modes. The similarities suggest that the interior of the RdRp in coronaviruses are attractive binding spot for small compounds in general.

The main enzymatic unit of Covid-19 RdRp, NSP12, mostly keeps its conformation between RNA free and RNA bound structures [12]. Figure 5 shows that cofactors NSP7 and NSP8 seek patches with kinetically active residues on the surface of NSP12. However, they are also in contact with kinetically less active areas. This should be analyzed in the light of fact that SARS-COV-2 RdRp (NSP12) cannot perform its function without NSP7 and NSP8 [12]. The distribution of kinetically very active and kinetically dormant residues may be important for the overall stability of NSP12, and also act as stochastic oscillator/transformer that translates random fluctuations of solvent and proteins into a regular vibrations that produce a regular rhythm of translation (i.e. act as a regular clock/oscillator).

## Spike glycoproteins and their interactions

### ACE2 binding patterns to SARS and Covid-19 Spike glycoproteins

The analysis of the contact patterns between the ACE2 receptor and the spike glycoprotein receptor binding domains (RBD) in SARS (pdb id 6cs2 [61]) and SARS-COV-2 (pdb id 6m0j [7]) reveals a difference in the distribution of kinetically active residues important for binding between RBD and ACE2 (Figure 6). The conformationally stable SARS-RBD has a smaller number of kinetically active and adjoining residues in direct contact with ACE2 (Figure 6a-c), while kinetically active residues in Covid-19 RBD are directly oriented and are in contact with the active residues in ACE2 (Figure 6d-f). In SARS active residues are mostly perpendicular to the interfacial plane (compare the distributions of C-alpha atoms in Figs. 6a and 6d). That should make the binding affinity between the Covid-19-RBD and ACE2 receptor stronger than between the SARS-RBD and ACE2 receptor. In both cases the predicted residues are recognized via the fastest vibrational mode (see [35]). For the list of hot residues and predictions see Figure S6 in the Supplementary material and the list below it. For 6cs2 the expected number of targets was between 22 and 25%, and that corresponds to the first, fastest mode for SARS Spike glycoprotein (chain B), and the fastest six modes for ACE2 (chain D). For 6m0j the expected number of targets was between 20 and 22%, and that corresponds to the first, fastest mode for Covid-19 Spike glycoprotein receptor binding domain (chain E), and fastest six modes for ACE2 (chain A).

### SARS-Cov spike glycoprotein and glycans

The analysis of kinetically active and adjoining residues in the SARS-Cov spike glycoprotein monomer (pdb id 6nb6) reveals that they are attractive binding spots for glycans (Figure 7). Glycans form the glycan shield, which was already suggested to assist in immune evasion similarly to the HIV-1 envelope trimer [63]. The kinetically active residues recognized by the SAGNM algorithm [35] can be used as target areas for drugs aimed at removing/disrupting the viral glycan shield. Those residues are not particularly hydrophobic and should be targeted by drugs that bind to hydrophilic patches, and electrostatically complementary.

### SARS Spike Glycoprotein RBD and human antibody fragment

We also analyzed the kinetically active residues in the structure formed by the SARS Spike Glycoprotein RBD and human neutralizing S230 antibody FAB fragment (pdb id 6nb6). The analysis reveals that S230 antibody binds to kinetically active residues in SRAS RBD, while heavy and light chains in S230 communicate via kinetically active residues (see Figure 8). The binding residues are mostly neutral to hydrophilic, thus any potential drug should be able to bind to similar surfaces (neutral/hydrophilic and stable).

### Boceprevir

With the expected number of targets between 15 and 20% of the total number of residues, which corresponds to the fastest normal mode, the SAGNM algorithm recognized the residues Val-20, Asn-28 and Cys-38 as kinetically hot in the SARS-Cov-2 main protease (pdb id 6wnp). That corresponds to the predictions Cys-16, Met-17, Val-18, Gln-19, Val-20, Thr-21, Cys-22, Gly-23, Thr-24, Thr-25, Thr-26, Leu-27, Asn-28, Gly-29, Leu-30, Trp-31, Leu-32, Asp-34, Val-35, Val-36, Tyr-37, Cys-38, Pro-39, Arg-40, His-41, Val-42, Phe-66, Leu-67, Val-68, Gln-69, Val-86, Leu-87, Lys-88, Cys-117, Tyr-118, Asn-119, Gly-120, Gly-143, Ser-144, Cys-145, Gly-146, Ser-147 and Met-162. Of all the residues in contact with boceprevir, the SAGNM algorithm recognized the residues Thr-25, Thr-26, Leu-27, His-41, Gly-143, Ser-144 and Cys-145, see Figure 9. See also Figure S8 in the Supplementary material for the distribution of hot and predicted residues.

### Eflornithine

The drug α-difluoromethylornithine (DMFO, Eflornithine) prohibits binding of the natural non-coded amino acid ornithine to the active site on the surface of *Trypanosoma brucei* Ornithine Decarboxylase (ODC, pdb id 1njj [60]). The binding of this drug should follow the binding patterns of ornithine. Figure 9 shows that SAGNM algorithm accurately detects binding sites for both ornithine and G418 (Geneticin), an aminoglycoside antibiotic. In contrast to chloroquine, both compounds bind preferably to the hydrophilic sites on the surface of ODC (Figure 10b-d). If applied to treat Covid-19, the drug eflornithine should bind to similar sites on the surface of Covid-19 proteins (hydrophobic and kinetically active, i.e. stable).

With the expected number of targets between 10 and 15% of the total number of residues, which corresponds to the fastest normal mode, the SAGNM algorithm recognized the residues Asp-44, Ala-281 and Phe-284 of the chain A of 1njj as kinetically hot. The corresponding predictions are Thr-21, Phe-40, Phe-41, Val-42, Ala-43, Asp-44, Leu-45, Gly-46, Asp-47, Ile-48, Gly-240, Thr-241, Arg-277, Tyr-278, Tyr-279, Val-280, Ala-281, Ser-282, Ala-283, Phe-284, Thr-285, Leu-286, Ala-287, Val-288, Glu-384, Asp-385, Met-386, Gly-387, Ala-388, Tyr-407, Val-408, Val-409 and Ser-410. See Figure S9 in the Supplementary material for their distribution.

## Discussion and conclusions

Covid-19 is the first modern, severe global pandemic caused by a coronavirus, and there are no guarantees that it will be the last. Our society therefore needs not only to develop an effective and efficient treatment for the current disease, but also has to have a set of protocols to promptly address all future, similar pandemics. In this manuscript we presented our strategy to recognize potential drug binding residues in human and viral proteins. We analyzed four currently approved drugs (chloroquine, ivermectin, remdesivir, sososbuvir, boceprevir and eflornithine). Our results indicate that small, drug like compounds preferentially bind to kinetically active and adjoining residues, thus seeking stable residues characterized by fast normal modes with small amplitude of fluctuations [35]. Some drugs preferentially seek active patches that are hydrophobic (chloroquine, ivermectin), while others prefer hydrophilic surfaces (remdesivir, sofosbuvir, eflornithine). We can postulate that in water environment drugs binding to hydrophilic patches will be more stable, as their removal will lead toward the reduction in structural entropy, but a full account of this proposition will require calculations of binding free energy differences based on full atom molecular dynamics, using, for instance, steered molecular dynamics simulations (SMD) [64-66]. We can also propose that the drugs/small molecules that bind to deep pockets will be more stable, and thus more effective. Our algorithm accurately recognizes such pockets as binding spots for drugs (Figures 1a, 3 and 10), and small peptides (see, in particular, Figure 6a in [35]).

Multidrug cocktails are frequently used to treat viral diseases [67]. Our analysis shows that in designing antiviral drug cocktails, binding affinity between drugs and kinetically active (stable) sites should be combined with the information on their hydrophobic and hydrophilic properties in attempt to avoid binding competition, increase drug cocktail efficiency, and reduce toxicity and other unwanted side effects.

In our analysis we used both viral-parasitic, as well as human proteins. The analysis shows that kinetically active residues exist in both human and non-human proteins/enzymes and that drugs bind indiscriminately to them regardless of their origin. The compounds that bind to human proteins potentially offer longer lasting treatments as host cells and tissues have less chance of developing drug resistance through single point mutations.

The procedure we described here is fast and effective, and can analyze a protein structure much faster than computationally demanding molecular dynamics simulations. Its advantage is not in its efficiency, but also in its ability to suggest general binding patterns between proteins and drugs or small peptides. It can be used to filter binding areas on protein surfaces and thus facilitate preclinical stages in drug design. Binding spots in various proteins can be very effectively predicted with our SAGNM approach and accessed with other bioinformatics tools for charge and shape complementarity, binding affinity, atomic mass and other properties as well. However, the SAGNM algorithm cannot determine binding free energies, or binding orientations of small molecules. For that aim other docking tools or molecular dynamics should be applied as we explained above. Our results are concordant with full atom docking and simulations studies [13-19] that emphasized sofosbuvir, remdesivir, hydroxychloroquine and ivermectin, compounds that we also analyzed. This indicates that protein-ligand docking is a multistep process, guided both by coarse grained properties of a bigger binding partner, and detailed, atomic scale properties of binding pocket and a small ligand.

Recent advances in machine learning helped advance our ability to predict and design protein structures [68], but the full theoretical foundations for protein folding and binding is still lacking. The quality of the machine learning protocol directly depends on the quality and size of training data sets and thus in many ways follows classical methods based on statistical potentials and homology modeling [69]. Our results can also help in that respect as they offer interpretation on how residue packing inside protein segments guides their assemblage.

Our results depicted here show that in proteins that interact with small, drug-like molecules contacting scaffolds are surrounded by kinetically hot residues. Similar conclusion related to protein-protein interactions was given in [35], but, as we showed above, the full binding behavior is not accessible only through the analysis of kinetically active residues and their neighbors. The full atom analysis is still required for the detailed assessment of protein-drug binding. The coarse grained analysis (SAGNM algorithm) thus, perceives only the outline of the binding funnel, while the full atom analysis (binding free energy) sees finer patterns inside that outline. Their combination may offer an overall improvement in binding prediction. This approach should be in principle be similar to the current improvement in Deep Neural Networks (DNN) architectures aimed at image recognition and classification by Brendel and Bethge [70]. The improvement is based on the splitting images into small local image features (e.g. outlines) without taking into account their spatial ordering, a strategy closely related to the pre deep-learning bag-of-feature (BoF) models. This image classification improvement is based on the observation, that standard DNN architectures perceive images primarily through textures, as opposed to human perception, which is primarily based on the outlines and shapes of objects [71]. In our case, the outline is determined primarily by the protein, and fine binding features ("binding textures") stem from the joint properties of the smaller binding partner and the binding outline determined by the protein. In that sense the SAGNM approach is similar to human vision, and molecular docking and dynamics studies to the machine, DNN based, vision. This observation opens a space for further work, where molecular binding will be treated as a two-step process where the coarse-grained shape of binding funnel will be determined by the larger partner in the first step, and final binding position and orientation by the multiple and detailed features of the binding funnel and a smaller partner inside that funnel.

**Acknowledgement:** The author would like to thank G.A. Mansoori for his encouragements to undertake this activity and reading the manuscript.

| Drug | Indication | Dosage in individuals aged ≥ 12 years | Effectiveness | Side effects | Precautions in patients with complications | | | |
|---|---|---|---|---|---|---|---|---|
| | | | | | Cardio-Pulmonary | Renal | Hepatic | Retinal |
| Chloroquine | Treatment Prevention | 500-600 mg weekly | Malaria, Amebiasis, Porphyria Cutanea Tarda | Serious | Yes | Yes | Yes | Yes |
| Ivermectin | Treatment Prevention | 3-15 mg once | Parasitic infestations | Mild-Serious | No | Yes | Yes | No |
| Remdesivir | Treatment | 100-200 mg daily | Ebola, Marburg virus diseases | Mild | No | Yes | No | No |
| Sofosbuvir | Treatment | 400 mg daily | Hepatitis-C, HIV | Mild - Moderate | Yes | Yes | Yes | Yes |
| Boceprevir | Treatment | 200 mg daily | Hepatitis-C | Mild-Serious | Yes | No | Yes | Yes |
| α-Difluoromethylornithine | Treatment | 300-400 mg/kg/day, cream | Trypanosomiasis, reduction of facial hair in women | Mild-Serious | No | No | Yes | No |

**Table 1.** Comparison of existing drugs currently being tested for the antiviral treatment and prevention of Covid-19 through drug repurposing.

# Figures

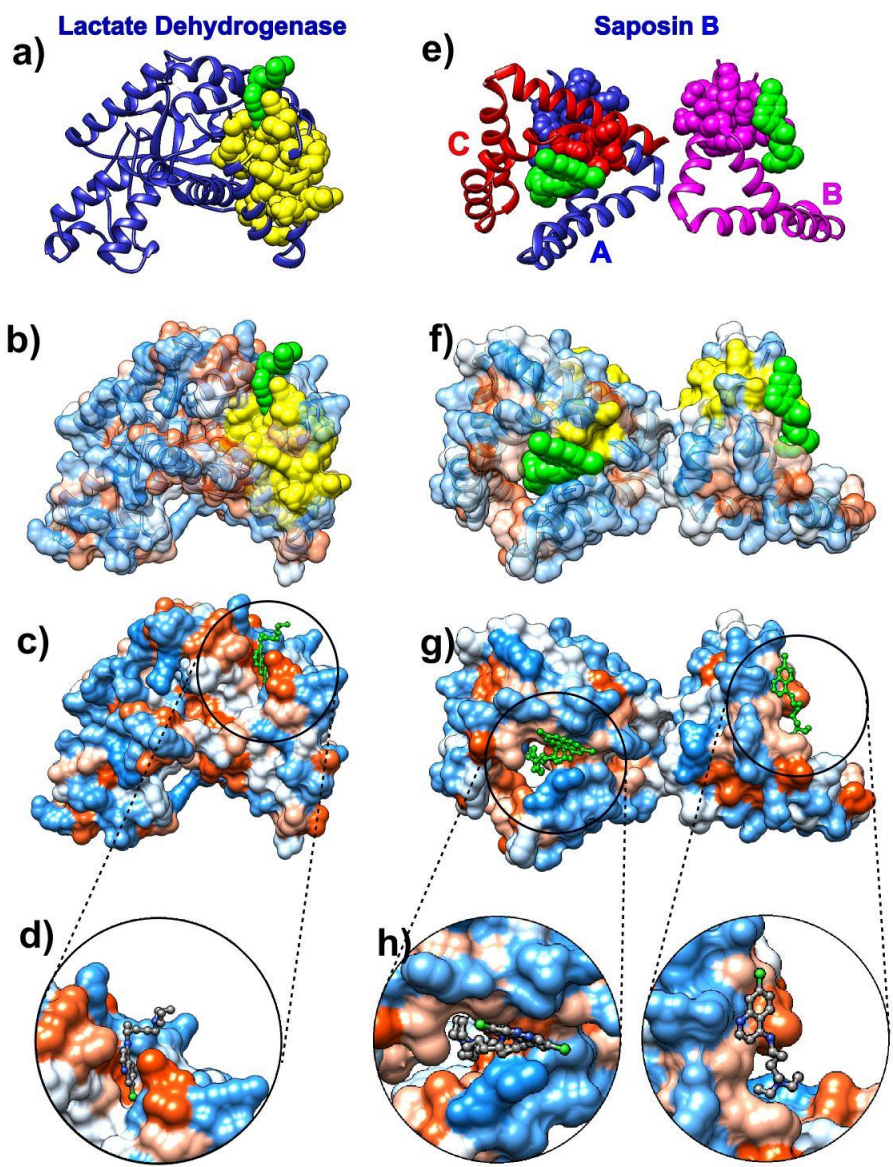

**Figure 1.** Chloroquine and its target proteins. Images on the left depict chloroquine bound to cofactor binding site of *Plasmodium Falciparum* Lactate Dehydrogenase (pdb id 1cet). Images on the right depict chloroquine bound to Saposin B (pdb id 4v2o). a) Lactate Dehydrogenase is depicted as blue ribbon, SAGNM predictions are yellow and Chloroquine green atoms. b) Lactate Dehydrogenase is depicted as transparent hydrophobic surface (chain is visible as ribbon inside surface). SAGNM predictions are depicted as yellow atoms and Chloroquine as green atoms. c) Lactate Dehydrogenase is depicted as opaque hydrophobic surface and chloroquine as green balls and sticks. d) The inset shows Chloroquine within the hydrophobic pocket. e) Saposin B chains depicted as blue (chain A), pink (chain B) and red (chain C) ribbons. SAGNM predictions are depicted as blue, pink and red atoms. Chloroquine molecules are shown as green atoms. f) Saposin B is depicted as transparent hydrophobic surface. SAGNM predictions are depicted as yellow atoms and Chloroquine green atoms. g) Saposin B is depicted as opaque hydrophobic surface and Chloroquine as green balls and sticks. h) The two insets show Chloroquine molecules within the hydrophobic pockets on the surface of the Saposin B trimer. The figure is produced with the UCSF Chimera program [36].

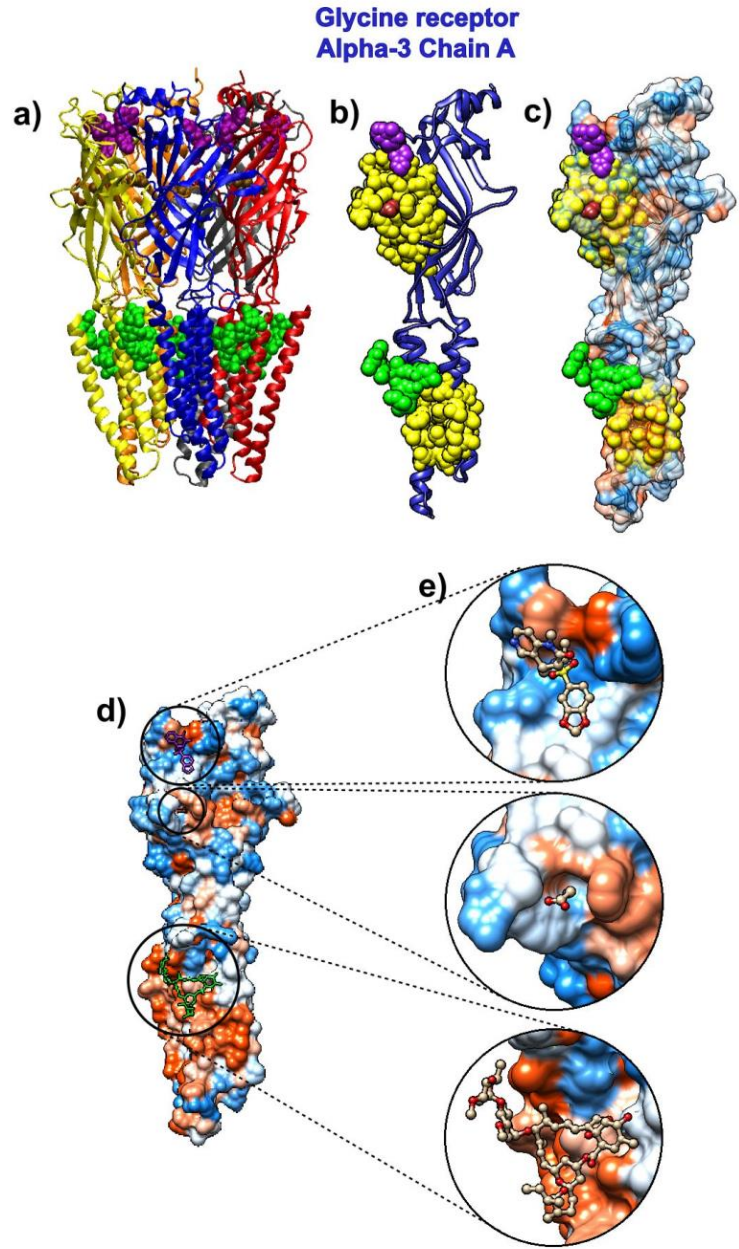

**Figure 2.** Ivermectin and its target protein, Human glycine receptor Alpha-3 (pdb id 5vdh). a) Five pentamer chains (A to R) represented as ribbons. Glycine molecules are brown and represented as atoms. 7C6 molecules are represented as purple atoms. Ivermectin is represented via green atoms. b) Chain A from Human glycine receptor Alpha-3 represented as blue ribbon. SAGNM predictions are depicted as yellow atoms. Glycine molecule is brown and represented vias atoms. 7C6 molecule is represented as purple atoms. Ivermectin is represented via green atoms. c) Chain A from Human glycine receptor Alpha-3 depicted as transparent hydrophobic surface. SAGNM predictions are yellow, glycine molecule are brown atoms, 7C6 molecule is represented as purple atoms and Ivermectin as green atoms. d) Chain A from Human glycine receptor Alpha-3 depicted as opaque hydrophobic surface. Glycine molecule are brown balls and sticks, 7C6 molecule is represented as purple and Ivermectin as green balls and sticks. e) The three insets show Glycine, 7C6 and ivermectin molecules inside the hydrophobic pockets on the surface of the chain A of Human glycine receptor Alpha-3. The figure is produced with the VMD and UCSF Chimera programs [36, 37].
17

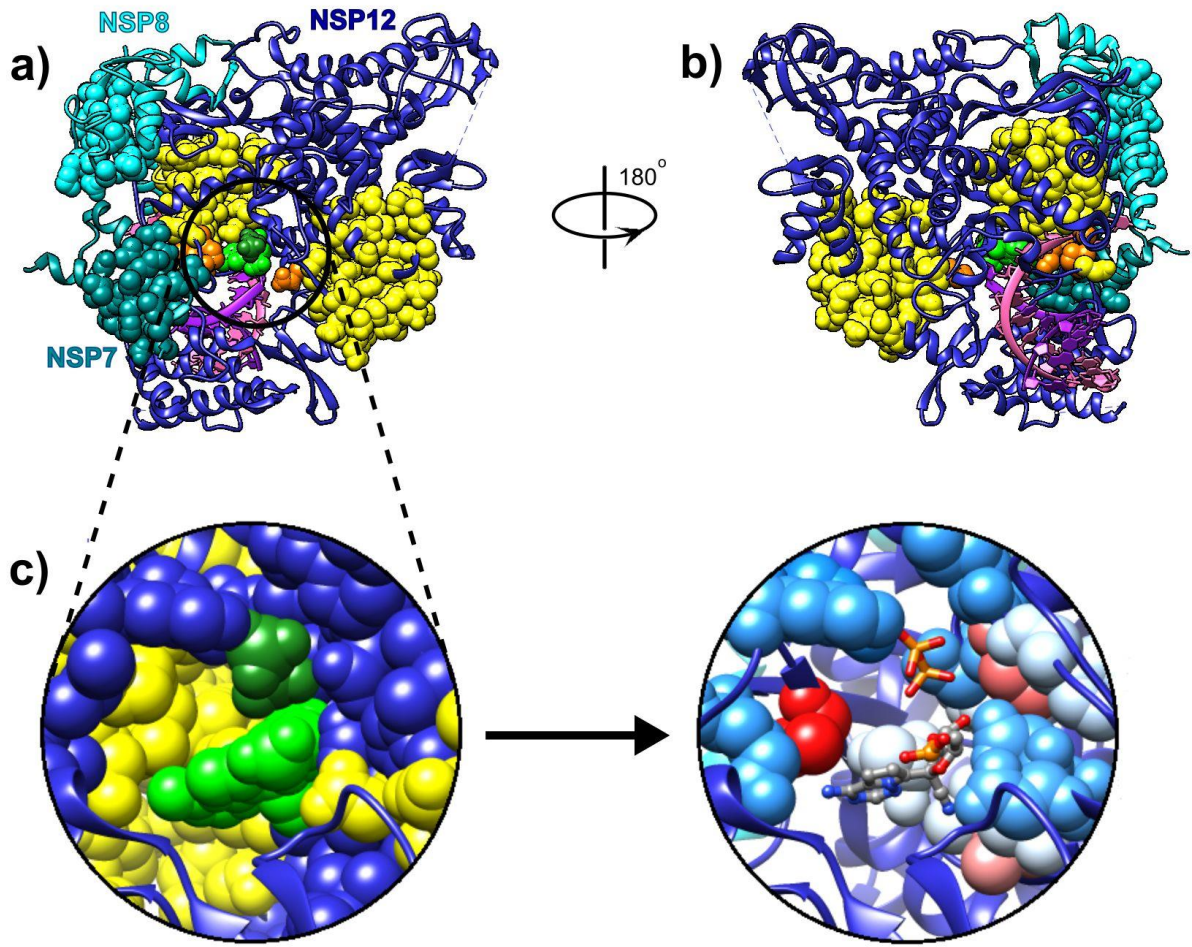

**Figure 3.** Remdesivir bound to the primer RNA inside the central channel of SARS-COV-2 RNA dependent RNA polymerase (RdRp), NSP12) (pdb id 7bv2 described in [12]). a) Three RNA polymerase chains, NSP 12, NSP7 and NSP8, represented as blue, cyan, and dark cyan ribbons. Remdesivir is represented via green atoms, and pyrophosphate as dark green atoms. The dashed lines represent protein segments missing from the deposited structure. b) The same structure rotated approximately 180º around the vertical axis. c) Remdesivir and pyrophosphate inside the binding pocket, surrounded by the yellow SAGNM predictions (Left), and inside the pocket with contact residues colored by hydrophobicity.

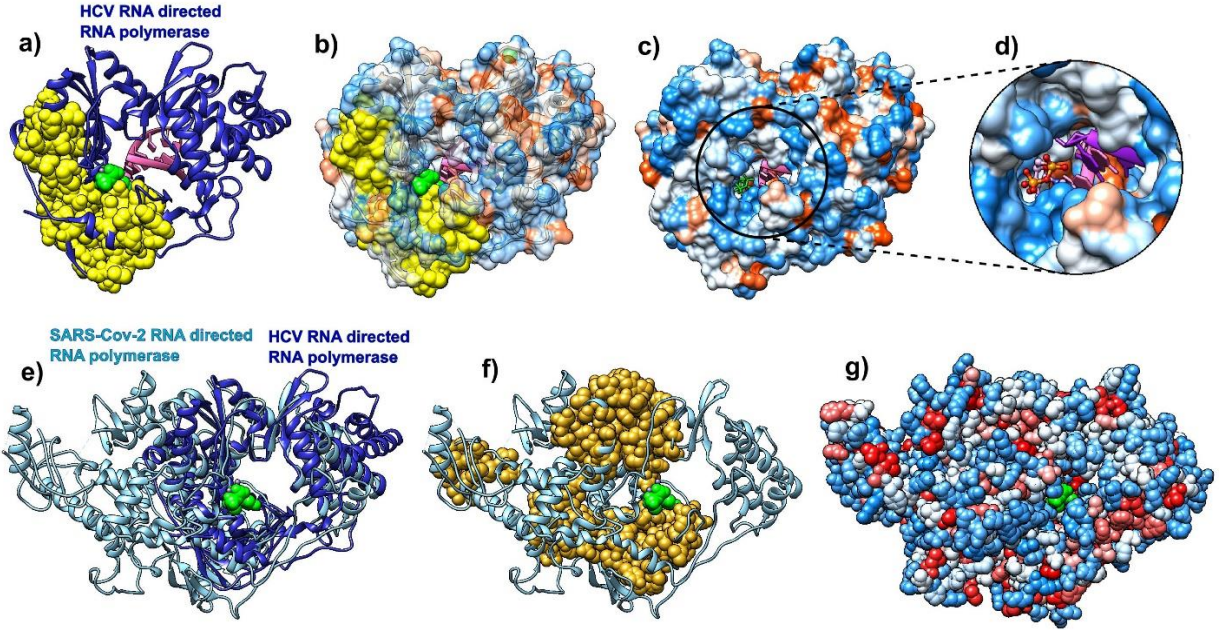

**Figure 4.** Comparative analysis of HCV (pdb id 4wtg, chain A, left) and Covid-19 RNA directed RNA polymerase (RdRp, pdb id 6m71, chain A, right). a) HCV RNA directed RNA polymerase is depicted as blue ribbon, RNA is purple, and sofosbuvir is green molecule (full atom representation). b) HCV RdRp is represented as a transparent hydrophobic surface, SAGNM predictions are yellow and Sofosbuvir is represented via green atoms. c) HCV RdRp is represented as an opaque hydrophobic surface and Sofosbuvir is represented via green sticks. d) The inset shows Sofosbuvir inside the polymerase catalytic core. e) HCV RdRp (blue ribbon) structurally aligned with Covid-19 RdRp (light blue ribbon). Sofosbuvir is green molecule inside the HCV RdRp catalytic core. f) Covid-19 RdRp as light blue ribbon. SAGNM predictions are dark yellow atoms. Sofosbuvir is green molecule inside the catalytic core. The position stems from the structurally aligned HCV RdRp. g) Covid-19 RdRp as hydrophobically colored atoms (residues hydrophobicities). Sofosbuvir is green molecule inside the catalytic core. The position stems from the structurally aligned HCV RdRp. With Covid-19 RNA polymerase sofosbuvir's position corresponds to the position it has when bound to HCV RNA polymerase.

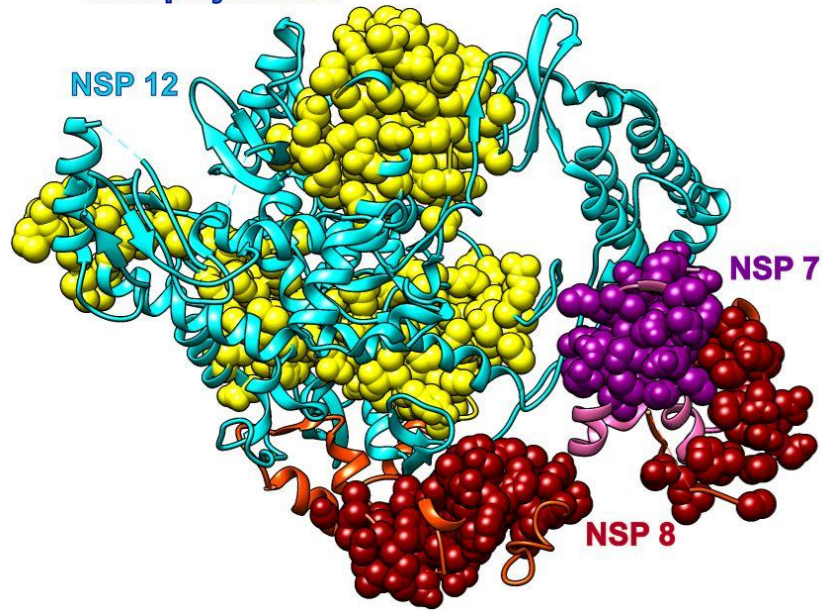

**Figure 5.** Covid-19 RNA directed RNA polymerase with cofactors NSP7 and NSP8 (pdb id 6m71). The NSP 12 chain is cyan, and its SAGNM predictions are yellow. The NSP 7 chain is pink and its SAGNM predictions are purple. The NSP 8 chain is orange and it SAGNM predictions are dark red. The dashed lines represent segments missing from the coordinates file.

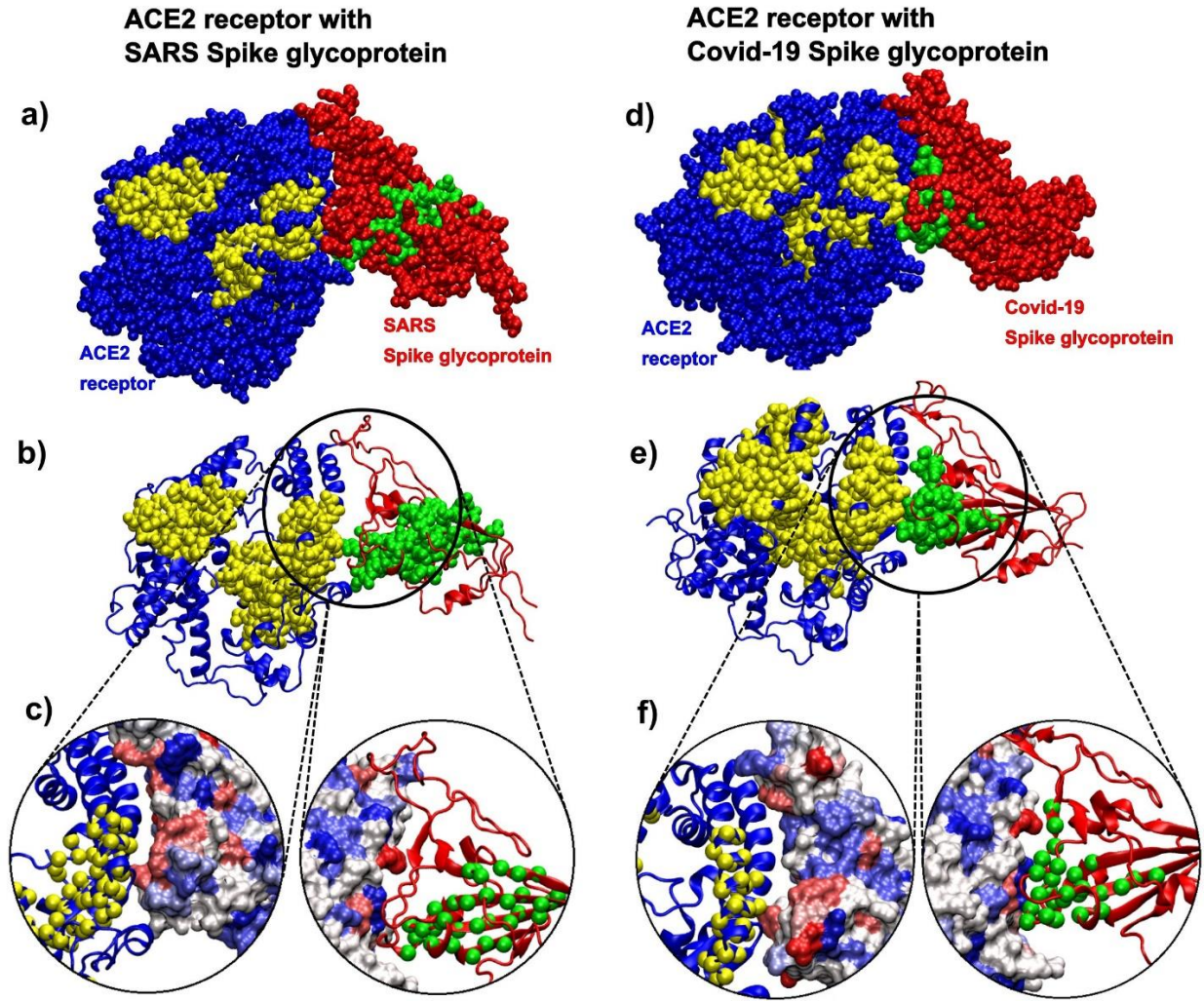

**Figure 6.** SARS Spike glycoprotein chain B RBD and ACE2 receptor (pdb id 6cs2) in comparison to Covid-19 Spike glycoprotein chain A RBD and ACE2 receptor (pdb id 6m0j). a) ACE2 receptor is represented via blue atoms and its SAGM predictions are yellow atoms. SARS Spike glycoprotein is represented via red atoms, and its SAGNM predictions and green atoms. b) ACE2 receptor is represented as a blue ribbon and its SAGM predictions are yellow atoms. SARS Spike glycoprotein is the red ribbon, and its SAGNM predictions and green atoms. c) Contact areas for both chains represented as hydrophobicity surfaces. The contact chains in each case are shown as ribbons, and predictions are represented via C-alpha atoms only. d) ACE2 receptor is represented via blue atoms and its SAGM predictions are yellow atoms. Covid-19 Spike glycoprotein is represented via red atoms, and its SAGNM predictions and green atoms. e) ACE2 receptor is represented as a blue ribbon and its SAGM predictions are yellow atoms. Covid-19 Spike glycoprotein is the red ribbon, and its SAGNM predictions and green atoms. f) Contact areas for both chains represented as hydrophobic surfaces. The contact chains in each case are shown as ribbons, and predictions are represented via C-alpha atoms only.

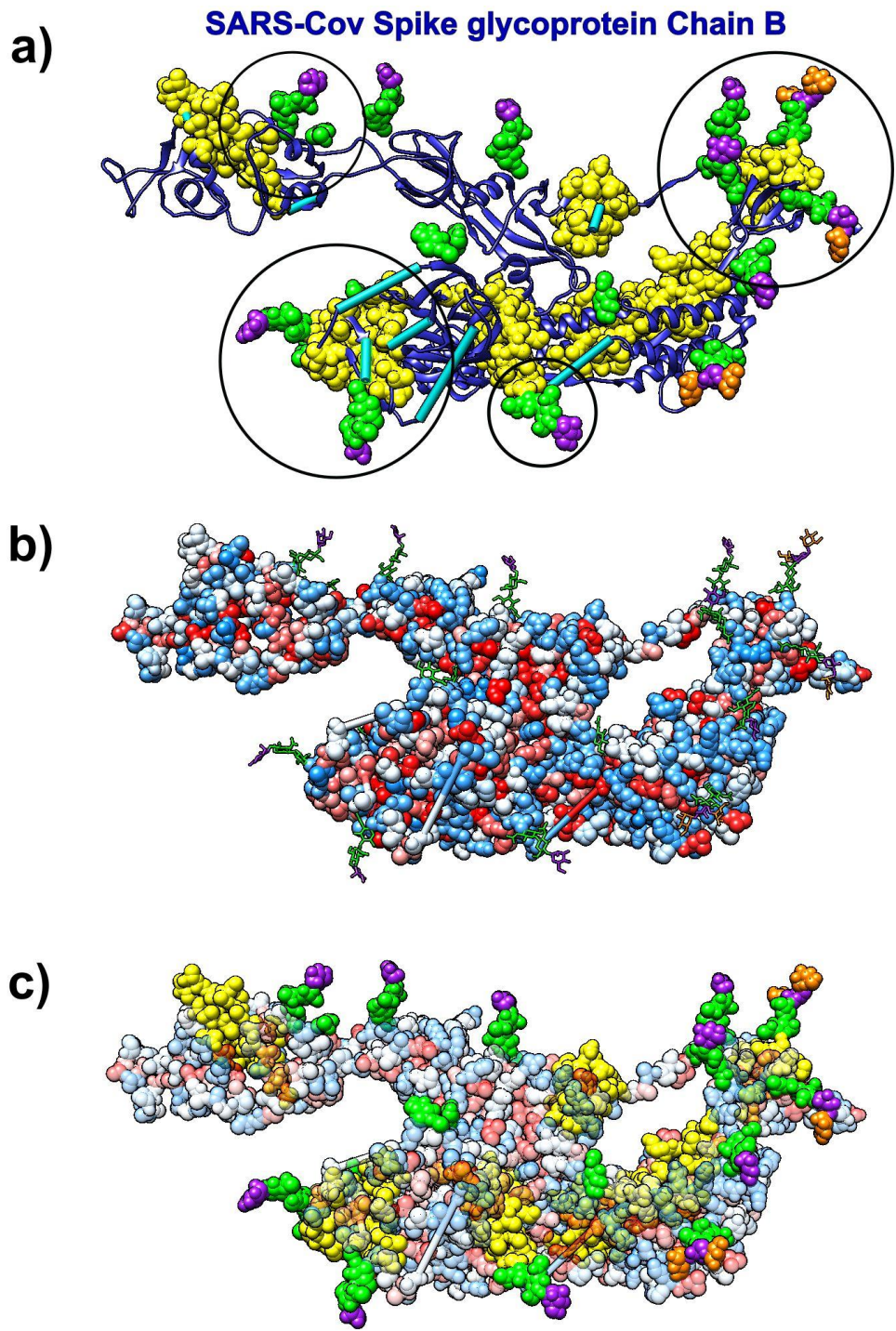

**Figure 7.** SARS-Cov spike glycoprotein (Chain B, pdb id 6nb6) with glycans (NAG, BMA, MAN) bound to it. a) Ribbon like representation of SARS spike glycoprotein. The SAGNM predictions are yellow atoms. BMA molecules are represented via purple atoms. MAN molecules are represented via orange atoms. NAG molecules are represented via green atoms. Cyan bars represent missing glycoprotein segments. b) SARS spike glycoprotein depicted via hydrophobicity colored atoms. Glycans (NAG, BMA, MAN) are represented via colored bonds (same colors as above). c) SARS spike glycoprotein depicted via transparent hydrophobicity colored atoms. Glycans (NAG, BMA, MAN) are represented via colored bonds (same colors as above). The SAGNM predictions are yellow atoms. Glycans (NAG, BMA, MAN) are represented via colored atoms.

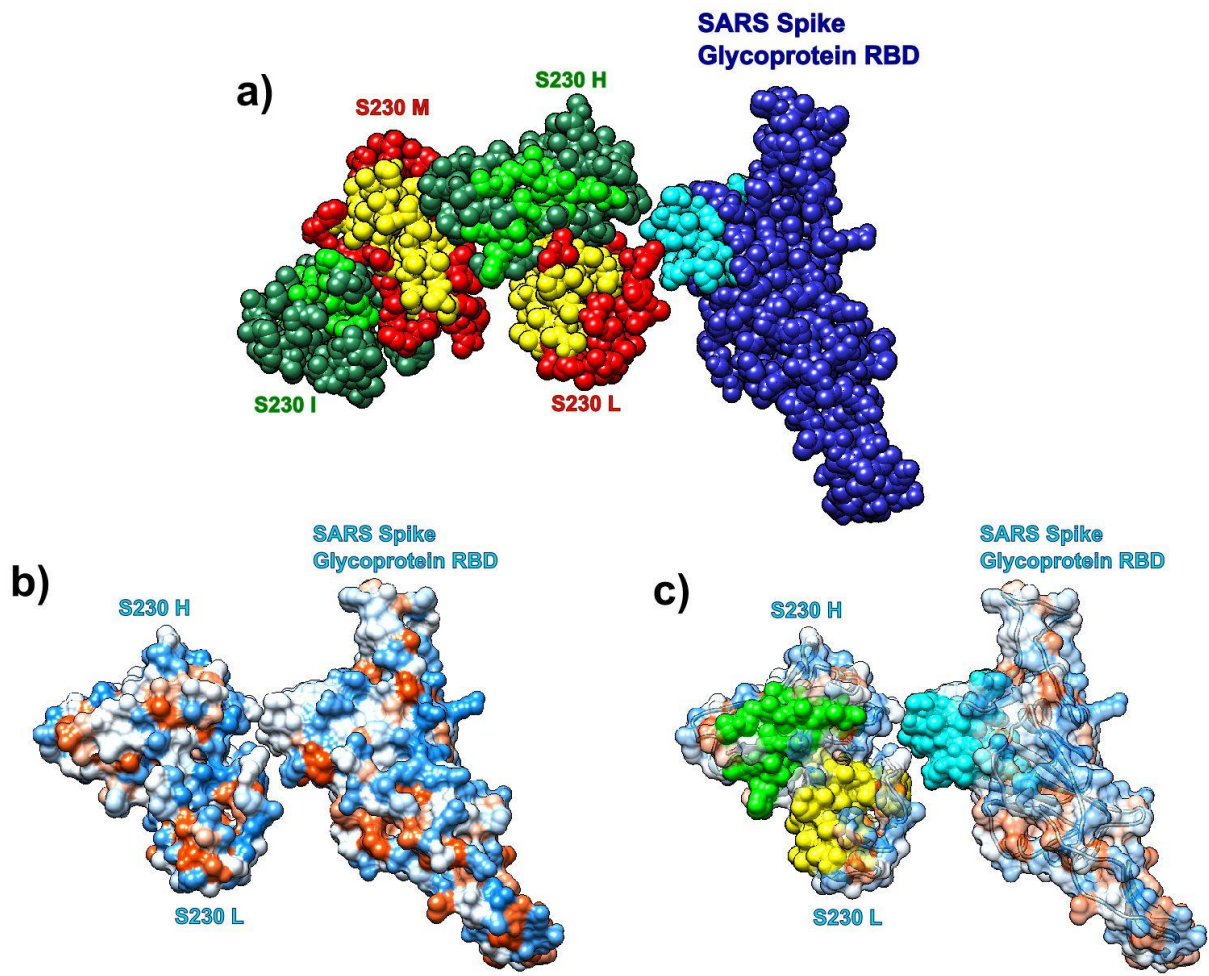

**Figure 8.** Receptor binding domain (RBD) of SARS-COV spike glycoprotein (Chain A, pdb id 6nb6) with human neutralizing S230 antibody FAB fragment. a) SARS-COV RBD (blue, chain A) with heavy (green, H and I) and light (red, L and M) chains. Predictions are cyan (SARS), yellow (S230 light) and light green (S230 heavy). b) Hydrophobic surface of SARS RBD bound to S230 (chains H and L). c) Transparent hydrophobic surface of SARS RBD and S230 (chains H and L) with predictions.

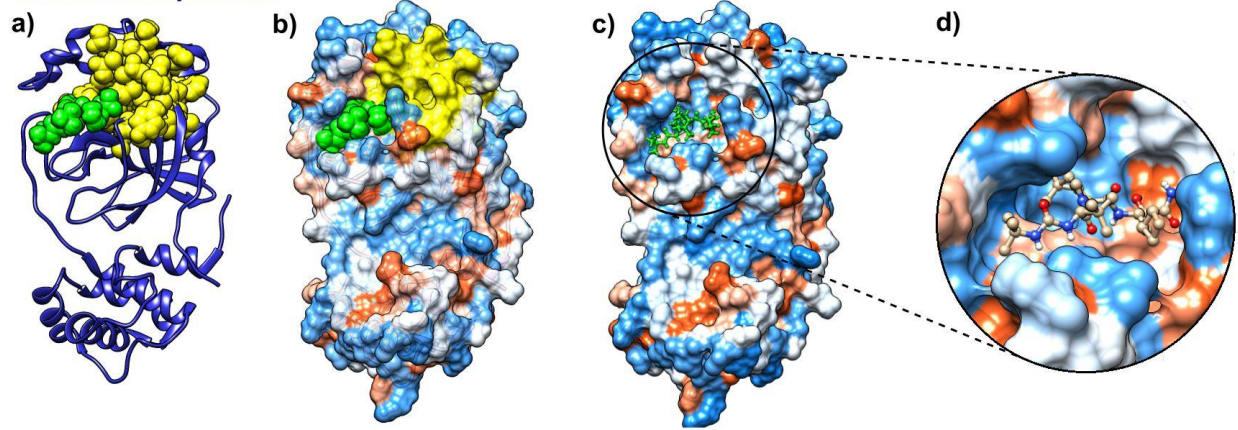

**Figure 9.** Boceprevir and its target protein Covid-19 (SARS-COV-2) main proease (pdb id 6wnp). a) Covid-19 Main protease is depicted as blue ribbon, SAGNM predictions are yellow and Boceprevir as green atoms. b) Covid-19 main protease is depicted as a transparent hydrophobic surface, SAGNM predictions are yellow and Chloroquine is green molecule. b) Covid-19 main protease is depicted as an opaque hydrophobic surface and Chloroquine is depicted via green balls and sticks. d) The inset shows Boceprevir inside the binding pocket.

# Ornithine decarboxylase

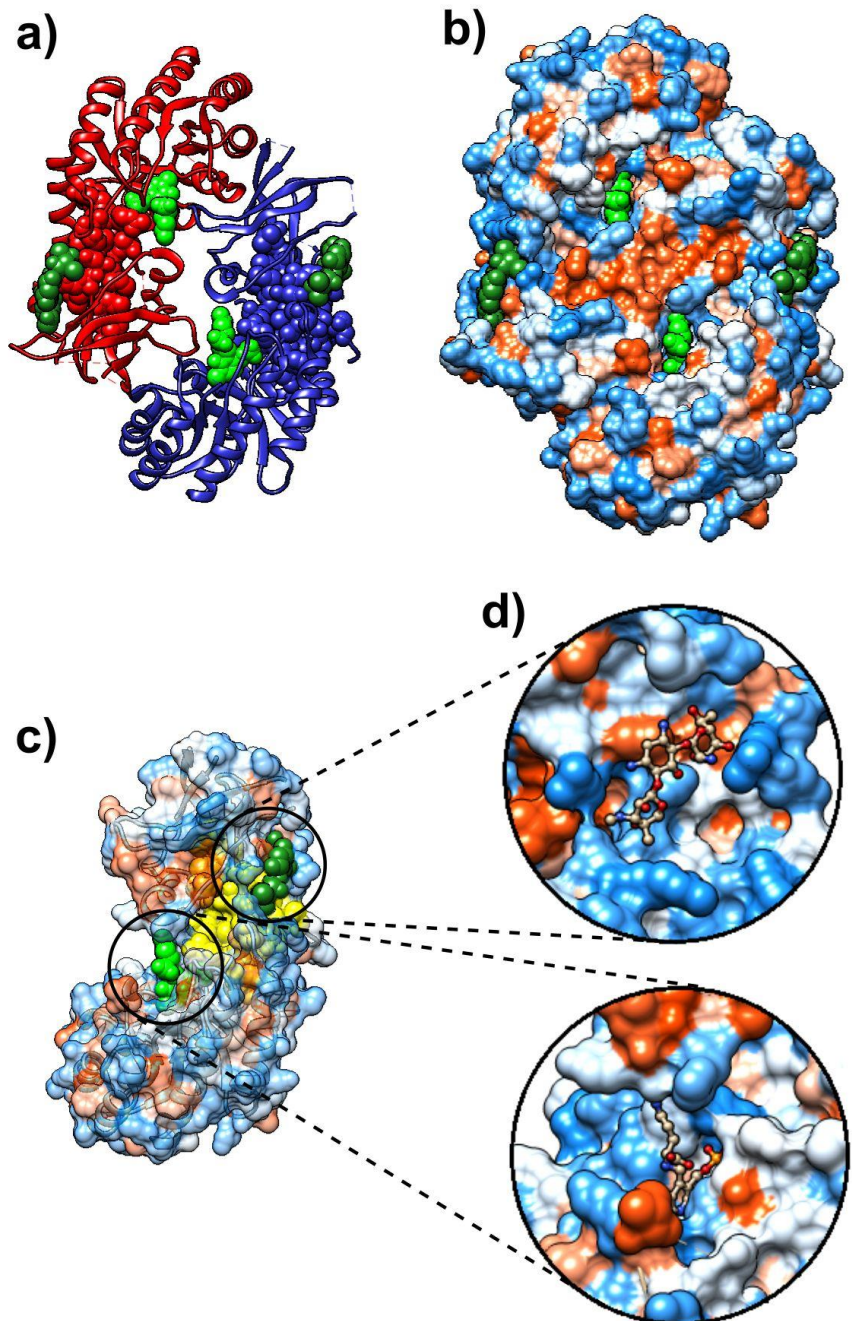

**Figure 10.** D-ornithine and its target protein *Trypanosoma brucei* ornithine decarboxylase (pdb id 1njj). a) Ornithine decarboxylase chains A (red) and B (blue) depicted as ribbons, with D-ornithine and G-418 as green and dark green molecules, respectively. b) Ornithine decarboxylase chains A and B depicted as hydrophobicity surface, with D-ornithine and G-418 as green and dark green molecules, respectively. c) Ornithine decarboxylase chain A depicted as transparent hydrophobicity surface, with SAGNM predictions are yellow atoms, and with D-ornithine and G-418 as green and dark green molecules. d) D-ornithine and G-418 molecules depicted as colored bonds&sticks, correspondingly to the atom type, inside pockets on the surface of ornithine decarboxylase.

# Recognition of potential Covid-19 drug treatments through the study of existing protein-drug and protein-protein structures: an analysis of kinetically active residues

## Supplementary material

Ognjen Perišić[1]
(1) Big Blue Genomics, Vojvode Brane 32, 11000 Belgrade, Serbia, ognjen.perisic@gmail.com

August 9, 2020

## Supplementary material figures

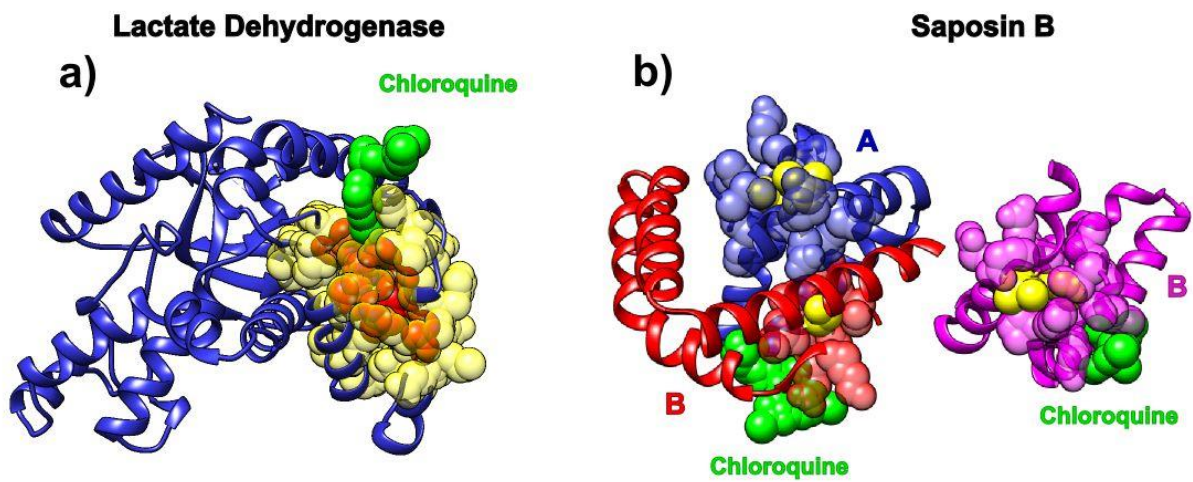

**Figure S1.** Hot residues and contact predictions for Lactate Dehydrogenase and Saposin B, determined by SAGNM. Images on the left depict chloroquine bound to cofactor binding site of *Plasmodium Falciparum* Lactate Dehydrogenase (pdb id 1cet). Images on the right depict chloroquine bound to Saposin B (pdb id 4v2o). a) Lactate Dehydrogenase is depicted as blue ribbon, SAGNM predictions are transparent yellow atoms and hot residues are red atoms. Chloroquine is depicted as green atoms. b) Saposin B chains are depicted as blue (chain A), pink (chain B) and red (chain C) ribbons. SAGNM predictions are depicted as transparent blue, pink and red atoms, and hot atoms are depicted as yellow atoms. Chloroquine molecules are depicted via green atoms.

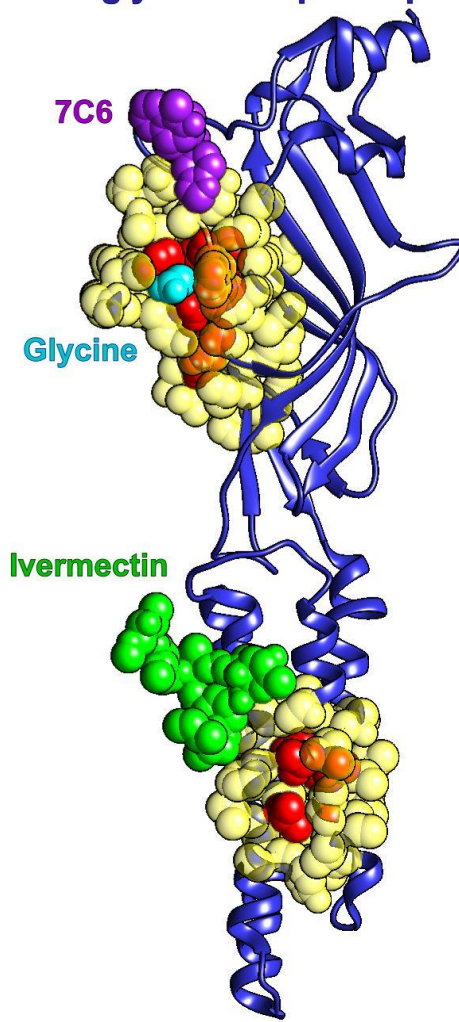

**Figure S2.** Ivermectin and its target protein, human glycine receptor Alpha-3 (pdb id 5vdh) with hot residues and contact predictions determined by SAGNM. Chain A from human glycine receptor Alpha-3 is represented as a blue ribbon. SAGNM predictions are depicted as transparent yellow atoms and hot residues are red atoms. Ivermectin is represented via green atoms. Glycine molecule is represented as cyan atoms. 7C6 molecule is represented as purple atoms.

The SAGNM algorithm recognized the residues Glu-157, Ser-158, Phe-168, Phe-207, Thr-208, Cys-209, Ile-210, Glu-211, Ser-238, Gly-256, Thr-259, Val-260, Val-294 and Leu-298 as from the chain A from 5vdh.pdb as kinetically hot. Their influence is spread to the residues Asn-38, Val-39, Thr-40, Cys-41, Pro-96, Asp-97, Leu-98, Phe-99, Phe-100, Ala-101, Ile-153, Met-154, Gln-155, Leu-156, Glu-157, Ser-158, Phe-159, Gly-160, Tyr-161, Thr-162, Met-163, Asn-164, Asp-165, Leu-166, Ile-167, Phe-168, Glu-169, Trp-170, Gln-171, Asp-172, Leu-195, Arg-196, Tyr-197, Cys-198, Thr-199, Lys-200, His-201, Asn-203, Thr-204, Gly-205, Lys-206, Phe-207, Thr-208, Cys-209, Ile-210, Glu-211, Val-212, Arg-213, Phe-214, His-215, Ile-234, Val-235, Ile-236, Leu-237, Ser-238, Trp-239, Val-240, Ser-241, Phe-242, Arg-252, Val-253, Ala-254, Leu-255, Gly-256, Ile-257, Thr-258, Thr-259, Val-260, Leu-261, Thr-262, Met-263, Thr-264, Cys-290, Leu-291, Leu-292, Phe-293, Val-294, Phe-295, Ser-296, Ala-297, Leu-298, Leu-299, Glu-300, Tyr-301 and Ala-302.

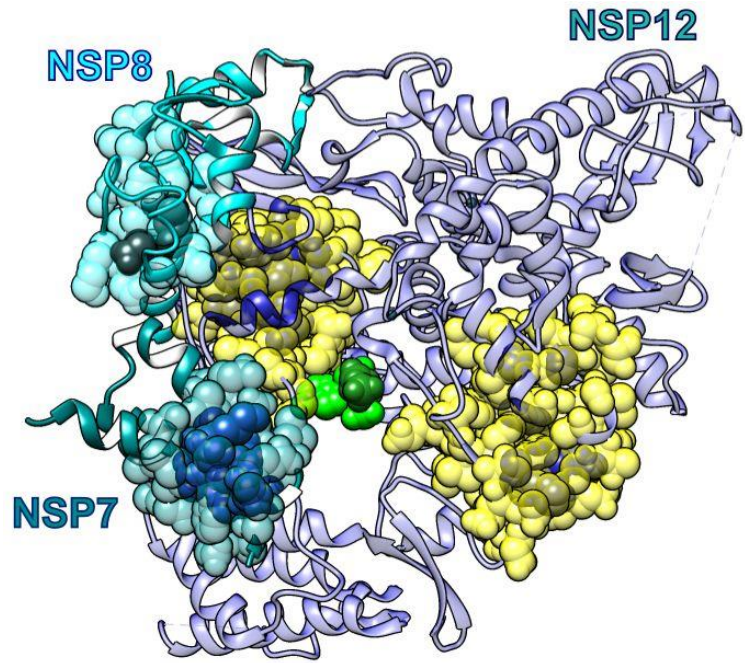

**Figure S3.** Remdesivir bound to the primer RNA inside the central channel of SARS-COV-2 RNA dependent RNA polymerase (RdRp), NSP12) (pdb id 7bv2 described in [12]). a) Three RNA polymerase chains, NSP 12, NSP7 and NSP8, represented as blue, cyan, and dark cyan ribbons. Remdesivir is represented via green atoms, and pyrophosphate as dark green atoms. SAGNM predictions for NSP12 are depicted as transparent yellow atoms and hot residues are blue atoms. SAGNM predictions for NSP7 are transparent green atoms and hot residues are blue atoms. SAGNM predictions for NSP9 are transparent cyan atoms and hot residues are gray atoms. Remdesivir is represented via green atoms, and pyrophosphate as dark green atoms.

The SAGNM algorithm recognized the residues Gly-503, Thr-538, Ile-539, Thr-540, Gln-541, Ala-558, Val-560, Ser-561, Val-609, His-613, Glu-665, Met-666, Val-667, Met-668, Ala-702, Ala-706, Phe-753, Cys-765 and Asn-767 in chain A as hot, the residues Asp-161 and Ile-185 in chain B as hot, and the residues Lys-7, Ser-10, His-36, Ile-39, Ala-48 and Lys-51 in chain C as hot. The influence of the hot residues in chain A is spread to the residues Ala-376, Asp-377, Pro-378, Asp-499, Lys-500, Ser-501, Ala-502, Gly-503, Phe-504, Pro-505, Phe-506, Asn-507, Asn-534, Val-535, Ile-536, Pro-537, Thr-538, Ile-539, Thr-540, Gln-541, Met-542, Asn-543, Leu-544, Lys-545, Ala-554, Arg-555, Thr-556, Val-557, Ala-558, Gly-559, Val-560, Ser-561, Ile-562, Cys-563, Ser-564, Thr-565, Val-605, Tyr-606, Ser-607, Asp-608, Val-609, Glu-610, Asn-611, Pro-612, His-613, Leu-614, Met-615, Gly-616, Trp-617, Cys-659, Ala-660, Gln-661, Val-662, Leu-663, Ser-664, Glu-665, Met-666, Val-667, Met-668, Cys-669, Gly-670, Gly-671, Ser-672, Leu-673, Tyr-674, Val-675, Lys-676, Ser-681, Ser-682, Gly-683, Asp-684, Gln-698, Ala-699, Val-700, Thr-701, Ala-702, Asn-703, Val-704, Asn-705, Ala-706, Leu-707, Leu-708, Ser-709, Thr-710, Tyr-748, Leu-749, Arg-750, Lys-751, His-752, Phe-753, Ser-754, Met-755, Met-756, Ile-757, Asp-761, Ala-762, Val-763, Val-764, Cys-765, Phe-766, Asn-767, Ser-768, Thr-769, Tyr-770, Ala-771, Gly-774, Leu-775, Val-776, Ala-777, Ser-778, Asn-781, Phe-782, Val-785, Thr-801, Glu-802, Thr-803, Asp-804 and Leu-805. The influence of the hot residues in chain B is spread to the residues Val-130, Val-131, Gln-157, Gln-158, Val-159, Val-160, Asp-161, Ala-162, Asp-163, Ser-164, Lys-165, Ala-181, Trp-182, Pro-183, Leu-184, Ile-185, Val-186, Thr-187, Ala-188 and Leu-189. And the influence of the hot residues in chain C is spread to the residues Ser-4, Asp-5, Val-6, Lys-7, Cys-8, Thr-9, Ser-10, Val-11, Val-12, Leu-13, Val-33, Gln-34, Leu-35, His-36, Asn-37, Asp-38, Ile-39, Leu-40, Leu-41, Ala-42, Thr-45, Thr-46, Glu-47, Ala-48, Phe-49, Glu-50, Lys-51, Met-52, Val-53 and Ser-54.

# HCV RNA directed RNA polymerase

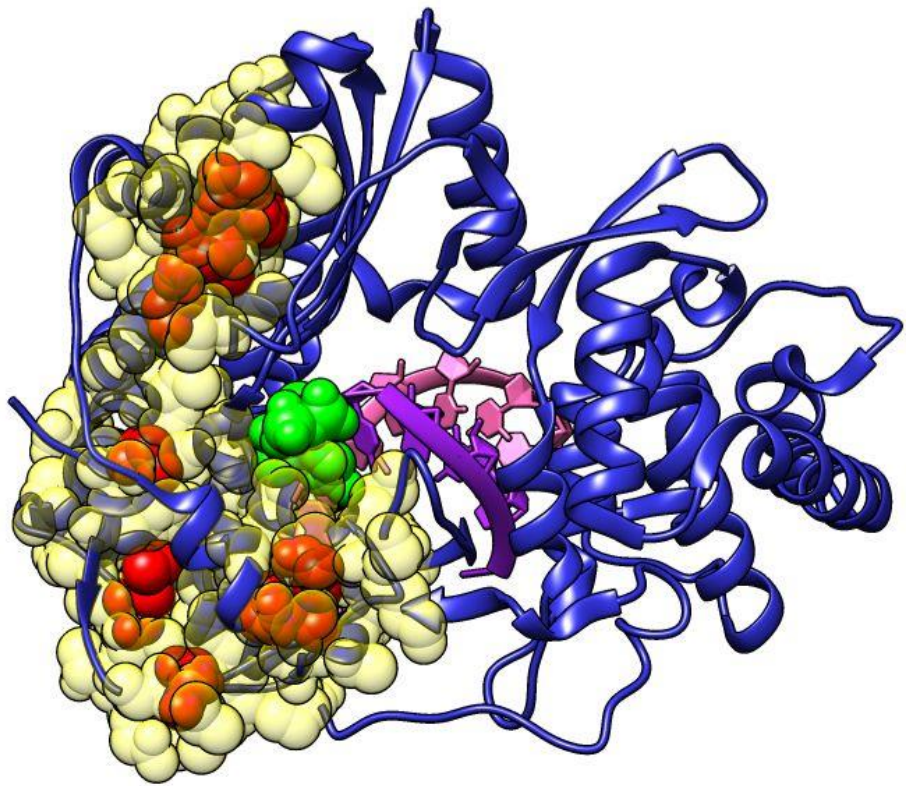

**Figure S4.** Sofosbuvir and its target protein Hepatitis C virus (HCV) RdRp (chain A in pdb id 4wtg) with hot residues and contact predictions determined by SAGNM. Chain A from HCV RdRp is represented as a blue ribbon. SAGNM predictions are depicted as transparent yellow atoms and kinetically hot residues are red atoms. Sofosbuvir is represented via green atoms. RNAs are pink and purple ribbons.

The SAGNM algorithm recognized the residues Ser-3, Tyr-4, Ala-9, Leu-10, Ile-11, Thr-12, Pro-13, Cys-14, Thr-40, Thr-41, Ser-42, Ser-44, Ala-45, Arg-48, Tyr-64, Pro-135, Thr-136, Thr-137, Ile-138, Met-139, Ala-140, Lys-141, Asn-142, Glu-143, Gly-153, Lys-154, Lys-155, Pro-156, Ala-157, Arg-158, Leu-159, Ile-160, Val-161, Asp-225, Ser-226, Thr-227, Val-228, Thr-229, Glu-230, Tyr-261, Val-262, Gly-263, Gly-264, Pro-265, Met-266, Phe-267, Asn-268, Ser-269, Lys-270, Gly-271, Gln-272, Thr-273, Cys-274, Gly-275, Tyr-276, Arg-277, Arg-278, Cys-279, Arg-280, Ala-281, Ser-282, Gly-283, Thr-294, Cys-295, Tyr-296, Val-297, Lys-298, Ala-299, Leu-300, Ala-301, Ala-302, Cys-303, Ala-305, Ala-306, Asn-335, Leu-336, Arg-337, Ala-338, Phe-339, Thr-340, Glu-341, Ala-342, Met-343, Thr-344, Arg-345, Tyr-346, Ser-347, Ala-348 and Pro-349 from 4wtg.pdb as predictions.

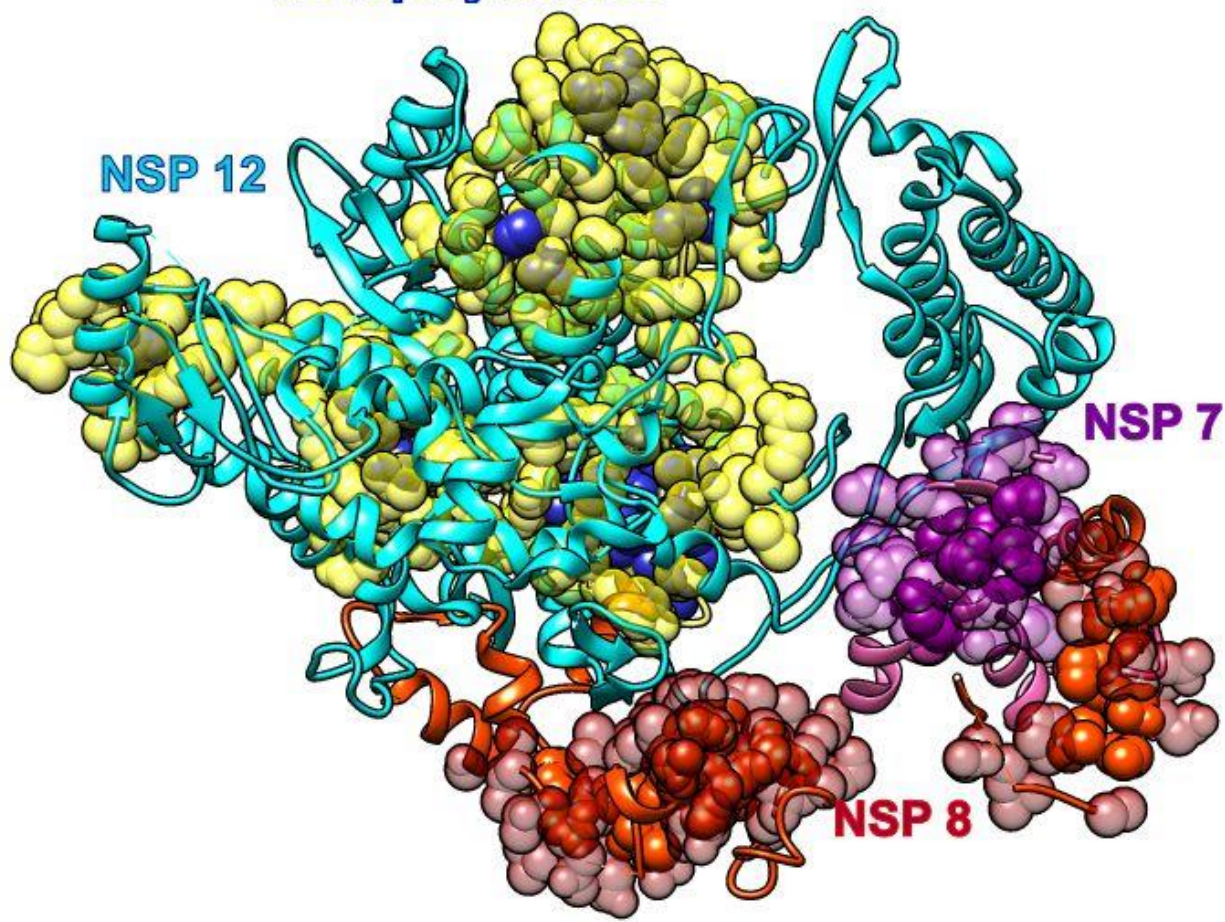

**Figure S5.** Covid-19 RNA directed RNA polymerase with cofactors NSP7 and NSP8 (pdb id 6m71). The NSP 12 chain is cyan, its SAGNM predictions are transparent yellow atoms, and hot residues are opaque blue atoms. The NSP 7 chain is pink, its SAGNM predictions are transparent purple atoms and hot residues are opaque purple atoms. The NSP 8 chain is orange, it SAGNM predictions are transparent red atoms and its hot atoms opaque orange atoms. The dashed lines represent segments missing from the coordinates file.

The SAGNM algorithm recognized the residues Gly-230, Cys-298, Cys-310, Asn-314, Gly-352, Ala-502, Gly-503, Thr-538, Ile-539, Thr-540, Gln-541, Gly-559, Val-560, Ser-561, Val-609, Glu-665, Val-667, Met-668, Ala-702, Ala-706, Val-763, Cys-765 and Asn-767 from the chain A from 6m71 as hot.

It also recognized the residues Met-196, Arg-197, Asn-198, Ala-199, Gly-200, Ile-201, Val-202, Thr-225, Thr-226, Pro-227, Gly-228, Ser-229, Gly-230, Val-231, Pro-232, Val-233, Val-234, Asp-274, Thr-276, Arg-279, Tyr-294, His-295, Pro-296, Asn-297, Cys-298, Val-299, Asn-300, Cys-301, Leu-302, Cys-306, Ile-307, Leu-308, His-309, Cys-310, Ala-311, Asn-312, Phe-313, Asn-314, Val-315, Leu-316, Phe-317, Ser-318, Gly-327, His-347, Phe-348, Arg-349, Glu-350, Leu-351, Gly-352, Val-353, Val-354, His-355, Asn-356, Ala-376, Asp-377, Pro-378, Leu-498, Asp-499, Lys-500, Ser-501, Ala-502, Gly-503, Phe-504, Pro-505, Phe-506, Asn-507, Gly-510, Lys-511,

Ala-512, Asn-534, Val-535, Ile-536, Pro-537, Thr-538, Ile-539, Thr-540, Gln-541, Met-542, Asn-543, Leu-544, Lys-545, Arg-555, Thr-556, Val-557, Ala-558, Gly-559, Val-560, Ser-561, Ile-562, Cys-563, Ser-564, Thr-565, Val-605, Tyr-606, Ser-607, Asp-608, Val-609, Glu-610, Asn-611, Pro-612, His-613, Leu-614, Met-615, Gly-616, Trp-617, Cys-659, Ala-660, Gln-661, Val-662, Leu-663, Ser-664, Glu-665, Met-666, Val-667, Met-668, Cys-669, Gly-670, Gly-671, Ser-672, Leu-673, Tyr-674, Val-675, Lys-676, Ser-681, Gly-683, Asp-684, Gln-698, Ala-699, Val-700, Thr-701, Ala-702, Asn-703, Val-704, Asn-705, Ala-706, Leu-707, Leu-708, Ser-709, Thr-710, Lys-751, His-752, Phe-753, Ser-754, Met-755, Met-756, Ile-757, Ser-759, Asp-760, Asp-761, Ala-762, Val-763, Val-764, Cys-765, Phe-766, Asn-767, Ser-768, Thr-769, Tyr-770, Ala-771, Gly-774, Leu-775, Val-776, Ala-777, Ser-778, Asn-781, Phe-782, Val-785, Pro-809 and Phe-812 as predictions.

The SAGNM algorithm recognized the residues Ile-132, Trp-154, Ile-156, Val-159, Asp-161, Leu-184, Ile-185, Val-186, Thr-187 and Ala-188 from the chain B from 6m71 as hot.
It also recognized the residues Lys-127, Leu-128, Met-129, Val-130, Val-131, Ile-132, Pro-133, Asp-134, Tyr-135, Asn-136, Thr-137, Tyr-138, Thr-146, Phe-147, Thr-148, Ala-150, Ser-151, Ala-152, Leu-153, Trp-154, Glu-155, Ile-156, Gln-157, Gln-158, Val-159, Val-160, Asp-161, Ala-162, Asp-163, Ser-164, Lys-165, Val-167, Leu-180, Ala-181, Trp-182, Pro-183, Leu-184, Ile-185, Val-186, Thr-187, Ala-188, Leu-189, Arg-190 and Ala-191 as predictions.

The SAGNM algorithm recognized the residues Val-6, Ser-10, Leu-35, Ile-39, Ala-48 and Met-52 from the chain C from 6m71 as hot.
It also recognized the residues Met-3, Ser-4, Asp-5, Val-6, Lys-7, Cys-8, Thr-9, Ser-10, Val-11, Val-12, Leu-13, Cys-32, Val-33, Gln-34, Leu-35, His-36, Asn-37, Asp-38, Ile-39, Leu-40, Leu-41, Ala-42, Thr-45, Thr-46, Glu-47, Ala-48, Phe-49, Glu-50, Lys-51, Met-52, Val-53, Ser-54 and Leu-55 as predictions.

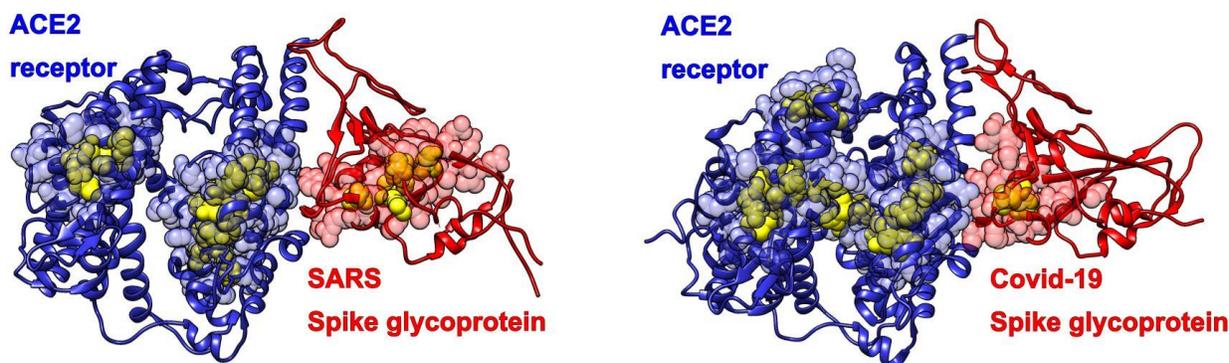

**Figure S6.** ACE2 receptor with SARS-COV Spike glycoprotein (pdb id 6cs2 left) and ACE2 receptor with Covid-19 Spike glycoprotein (6m0j.pdb, right). Predictions are blue transparent atoms (ACE2 receptor) and red transparent atoms (SARS and Covoid-19 Spike glycoprotein). Hot residues are yellow.

The SAGNM algorithm recognized the residues Asp-385, Phe-387, Pro-493, Arg-495 and Val-496 in the chain B of 6cs2, and predictions are residues Phe-334, Pro-335, Ser-336, Val-337, Trp-340, Glu-341, Arg-342, Asn-381, Val-382, Tyr-383, Ala-384, Asp-385, Ser-386, Phe-387, Val-388, Val-389, Lys-390, Gly-391, Leu-421, Ala-422, Trp-423, Thr-425, Arg-426, Asp-429, Ile-489, Gly-490, Tyr-491, Gln-492, Pro-493, Tyr-494, Arg-495, Val-496, Val-497, Val-498, Leu-499 and Ser-500.

The SAGNM algorithm recognized the residues Ser-43, Ala-46, Tyr-50, Met-62, Met-123, Val-172, Gly-173, Leu-176, Ala-348, His-373, His-374, Glu-375, His-378, Tyr-385, Ala-403, Gly-405, Ser-409 and Cys-498 as hot in the chain D of 6cs2.

The predictions are residues Leu-39, Phe-40, Tyr-41, Gln-42, Ser-43, Ser-44, Leu-45, Ala-46, Ser-47, Trp-48, Asn-49, Tyr-50, Asn-51, Thr-52, Asn-53, Ile-54, Thr-55, Asn-58, Val-59, Gln-60, Asn-61, Met-62, Asn-63, Asn-64, Ala-65, Gly-66, Trp-69, Ile-119, Leu-120, Asn-121, Thr-122, Met-123, Ser-124, Thr-125, Ile-126, Tyr-127, Gly-130, Lys-131, Trp-168, Arg-169, Ser-170, Glu-171, Val-172, Gly-173, Lys-174, Gln-175, Leu-176, Arg-177, Pro-178, Leu-179, Tyr-180, Phe-308, Ala-311, Ala-342, Cys-344, His-345, Pro-346, Thr-347, Ala-348, Trp-349, Asp-350, Leu-351, Gly-352, Phe-356, Arg-357, Ile-358, Leu-359, Phe-369, Leu-370, Thr-371, Ala-372, His-373, His-374, Glu-375, Met-376, Gly-377, His-378, Ile-379, Gln-380, Tyr-381, Asp-382, Met-383, Ala-384, Tyr-385, Ala-386, Ala-387, Gln-388, Pro-389, Leu-392, Arg-393, Asn-394, Gly-399, Phe-400, His-401, Glu-402, Ala-403, Val-404, Gly-405, Glu-406, Ile-407, Met-408, Ser-409, Leu-410, Ser-411, Ala-412, Ala-413, Met-474, Asp-494, Glu-495, Thr-496, Tyr-497, Cys-498, Asp-499, Pro-500, Ala-501, Ser-502, Ser-507, Asn-508, Arg-518, Thr-519, Tyr-521, Gln-522, Leu-558, Arg-559, Leu-560 and Gly-561.

The SAGNM algorithm recognized the residues Ser-43, Tyr-50, Met-62, Trp-69, Met-123, Val-172, Gly-173, Leu-176, Met-190, Ala-191, Asp-198, Ala-403, Gly-405, Ser-502, Arg-518, Leu-520 and Gln-522 of the chain A of 6m0j as hot.

The prediction are Leu-39, Phe-40, Tyr-41, Gln-42, Ser-43, Ser-44, Leu-45, Ala-46, Ser-47, Trp-48, Asn-49, Tyr-50, Asn-51, Thr-52, Asn-53, Ile-54, Thr-55, Asn-58, Val-59, Gln-60, Asn-61, Met-62, Asn-63, Asn-64, Ala-65, Gly-66, Asp-67, Lys-68, Trp-69, Ser-70, Ala-71, Phe-72, Leu-73, Gly-104, Ser-105, Val-107, Leu-108, Ile-119, Leu-120, Asn-121, Thr-122, Met-123, Ser-124, Thr-125, Ile-126, Tyr-127, Gly-130, Lys-131, Ser-167, Trp-168, Arg-169, Ser-170, Glu-171, Val-172, Gly-173, Lys-174, Gln-175, Leu-176, Arg-177, Pro-178, Leu-179, Tyr-180, Leu-186, Lys-187, Asn-188, Glu-189, Met-190, Ala-191, Arg-192, Ala-193, Asn-194, His-195, Tyr-196, Glu-197, Asp-198, Tyr-199, Gly-200, Asp-201, Tyr-202, Ala-342, His-373, His-374, Glu-375, Met-376, Gly-377, His-378, Glu-398, Gly-399, Phe-400, His-401, Glu-402, Ala-403, Val-404, Gly-405, Glu-406, Ile-407,

Met-408, Ser-409, Phe-464, Lys-465, Tyr-497, Cys-498, Asp-499, Pro-500, Ala-501, Ser-502, Leu-503, Phe-504, His-505, Val-506, Ser-507, Asn-508, Arg-514, Tyr-515, Tyr-516, Thr-517, Arg-518, Thr-519, Leu-520, Tyr-521, Gln-522, Phe-523, Gln-524, Phe-525, Gln-526, Met-579, Asn-580, Val-581, Pro-583 and Leu-584.

The SAGNM algorithm recognized the residues Phe-497 and Pro-507 of the chain E of 6m0j as hot.
The predictions are the residues Ile-402, Arg-403, Gly-404, Ser-438, Asn-439, Ser-443, Gly-447, Asn-448, Gln-493, Ser-494, Tyr-495, Gly-496, Phe-497, Gln-498, Pro-499, Thr-500, Asn-501, Val-503, Gly-504, Tyr-505, Gln-506, Pro-507, Tyr-508, Arg-509, Val-510 and Val-511.

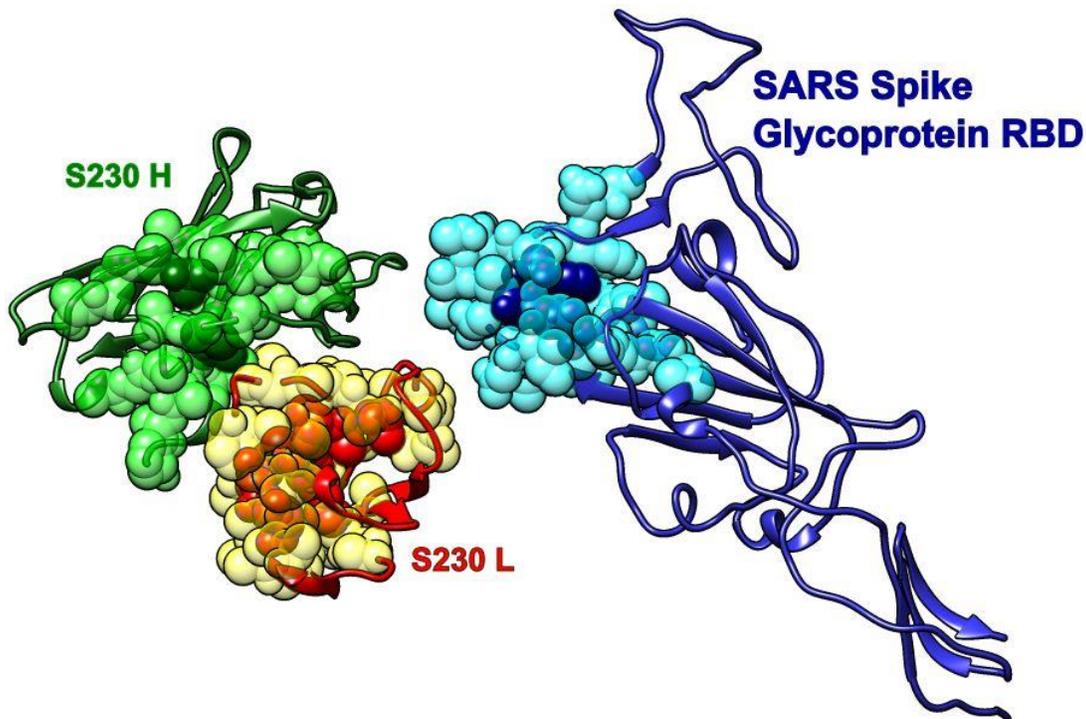

**Figure S7.** Receptor binding domain (RBD) of SARS-COV spike glycoprotein (Chain A, pdb id 6nb6) with human neutralizing S230 antibody FAB fragment. a) SARS-COV RBD (blue, chain A) with heavy (green, H and I) and light (red, L and M) chains. Predictions are cyan (SARS), yellow (S230 light) and light green (S230 heavy).

The SAGNM algorithm recognized the residues Ala-430, Phe-483 and Pro-493 from the RBD domain of SARS-COV spike glycoprotein (Chain A, pdb id 6nb6) as hot.
The predictions are Val-389, Lys-390, Gly-391, Thr-425, Arg-426, Asn-427, Ile-428, Asp-429, Ala-430, Thr-431, Ser-432, Thr-433, Gly-434, Asn-435, Asn-479, Asp-480, Tyr-481, Gly-482, Phe-483, Tyr-484, Thr-485, Thr-486, Thr-487, Ile-489, Gly-490, Tyr-491, Gln-492, Pro-493, Tyr-494, Arg-495, Val-496 and Val-497.
The SAGNM algorithm recognized the residues Glu-6, Val-37 and Tyr-95 from the chain H from 6nb6.pdb as hot. The corresponding predictions are Ala-2, Gln-3, Leu-4, Val-5, Glu-6, Ser-7, Gly-8, Gly-9, Ala-10, Ser-21, Cys-22, Ala-33, Met-34, His-35, Trp-36, Val-37, Arg-38, Gln-39, Ala-40, Pro-41, Gln-46, Trp-47, Leu-48, Thr-91, Ala-92, Val-93, Tyr-94, Tyr-95, Cys-96, Val-97, Thr-98, Gln-99, Gly-118, Gly-120 and Thr-121.
The SAGNM algorithm recognized the residues Glu-6, Val-37 and Tyr-95 from the chain I from 6nb6.pdb as hot. The corresponding predictions are Ala-2, Gln-3, Leu-4, Val-5, Glu-6, Ser-7, Gly-8, Gly-9, Ala-10, Ser-21, Cys-22, Ala-33, Met-34, His-35, Trp-36, Val-37, Arg-38, Gln-39, Ala-40, Pro-41, Gln-46, Trp-47, Leu-48, Thr-91, Ala-92, Val-93, Tyr-94, Tyr-95, Cys-96, Val-97, Thr-98, Gln-99, Gly-118, Gly-120 and Thr-121.
The SAGNM algorithm recognized the residues Gln-6, Cys-23, Trp-40, Phe-41, Ile-53, Tyr-92, Cys-93, Gly-104, Gly-106 and Thr-107 from the chain L from 6nb6.pdb as hot. The corresponding predictions are Val-3, Leu-4, Thr-5, Gln-6, Ser-7, Pro-8, Leu-9, Ser-10, Ala-19, Ser-20, Ile-21, Ser-22, Cys-23, Arg-24, Ser-25, Ser-26, Gln-27, Thr-36, Tyr-37, Leu-38, Asn-39, Trp-40, Phe-41, Gln-42, Gln-43, Arg-44, Pro-45, Pro-49, Arg-50, Arg-51, Leu-52, Ile-53, Tyr-54, Gln-55, Val-56, Ser-57, Asn-58, Arg-59, Phe-76, Val-88, Gly-89, Val-90, Tyr-91, Tyr-92, Cys-93, Met-94, Gln-95,

Gly-96, Ser-97, Pro-100, Pro-101, Thr-102, Phe-103, Gly-104, Gln-105, Gly-106, Thr-107, Lys-108, Val-109, Glu-110 and Ile-111.

The SAGNM algorithm recognized the residues Gln-6, Cys-23, Trp-40, Tyr-91, Tyr-92, Thr-107 and Val-109 from the chain M from 6nb6.pdb as hot. The corresponding predictions are Val-3, Leu-4, Thr-5, Gln-6, Ser-7, Pro-8, Leu-9, Ser-10, Leu-11, Pro-12, Ala-19, Ser-20, Ile-21, Ser-22, Cys-23, Arg-24, Ser-25, Ser-26, Gln-27, Thr-36, Tyr-37, Leu-38, Asn-39, Trp-40, Phe-41, Gln-42, Gln-43, Arg-44, Arg-51, Leu-52, Ile-53, Phe-76, Asp-87, Val-88, Gly-89, Val-90, Tyr-91, Tyr-92, Cys-93, Met-94, Gln-95, Gly-96, Gly-104, Gly-106, Thr-107, Lys-108, Val-109 and Glu-110.

In all cases the expected number of targets is between 10 and 15%. That corresponds to the first, fastest mode in each case.

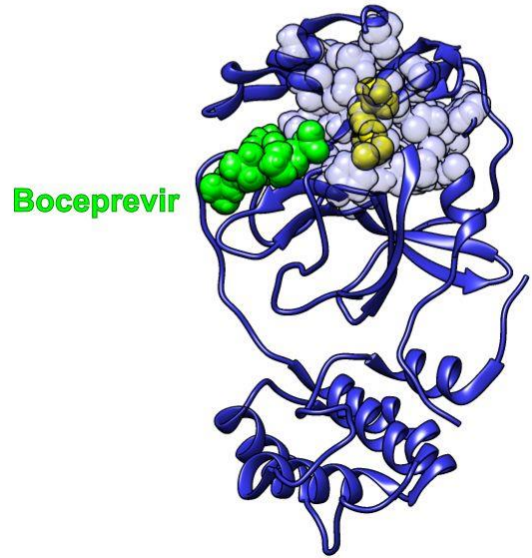

**Figure S8.** Hot residues and predictions for the SARS-Cov-2 main protease. The Covid-19 main protease (pdb id 6wnp) depicted as blue ribbon, with SAGNM predictions as transparent blue atoms, and hot residues as opaque yellow atoms. Boceprevir is depicted as a green molecule.

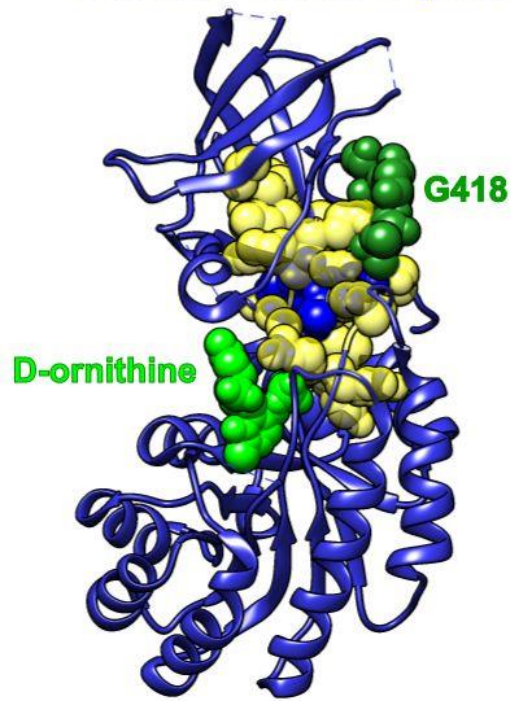

**Figure S9.** Ornithine decarboxylase (1njj.pdb) chain A depicted as blue ribbon, with SAGNM predictions as transparent yellow atoms, and hot residues as opaque blue atoms. D-ornithine and G-418 are depicted as green and dark green molecules, respectively.